\documentclass[aps,prb,twocolumn,amssymb,floatfix,epsfig]{revtex4}
\usepackage{graphicx,epsfig,psfrag,subfigure,amsopn}
\usepackage{color}
\usepackage{hyperref}
\usepackage{hypcap}
\usepackage{amssymb,amsbsy,amsmath}
\usepackage{amsmath} \textheight=22truecm
\newcommand\beq{\begin{equation}}
\newcommand\eeq{\end{equation}}
\newcommand\bea{\begin{eqnarray}}
\newcommand\eea{\end{eqnarray}}
\newcommand\bsq{\begin{subequations}}
\newcommand\esq{\end{subequations}}

\newcommand\al{\alpha}
\newcommand\ga{\gamma}
\newcommand\de{\delta}
\newcommand\De{\Delta}
\newcommand\ep{\epsilon}
\newcommand\si{\sigma}

\newcommand\lam{\lambda}
\newcommand\om{\omega}
\newcommand\ta{\theta}
\newcommand\dg{\dagger}
\newcommand\pa{\partial}
\newcommand\ua{\uparrow}
\newcommand\da{\downarrow}
\newcommand\la{\langle}
\newcommand\ra{\rangle}
\newcommand\non{\nonumber}

\newcommand\bib{\bibitem}

\begin{document}

\title{Josephson junctions of multiple superconducting wires}

\author{Oindrila Deb$^1$, K. Sengupta$^2$, and Diptiman Sen$^1$}

\affiliation{$^1$Centre for High Energy Physics, Indian Institute of Science,
Bengaluru 560012, India \\
$^2$Theoretical Physics Department, Indian Association for the Cultivation of
Science, Jadavpur, Kolkata 700032, India}

\date{\today}

\begin{abstract}

We study the spectrum of Andreev bound states and Josephson currents
across a junction of $N$ superconducting wires which may have $s$-
or $p$-wave pairing symmetries and develop a scattering matrix based
formalism which allows us to address transport across such
junctions. For $N \ge 3$, it is well known that Berry curvature
terms contribute to the Josephson currents; we chart out situations
where such terms can have relatively large effects. For a system of
three $s$- or three $p$-wave superconductors, we provide analytic
expressions for the Andreev bound state energies and study the
Josephson currents in response to a constant voltage applied across
one of the wires; we find that the integrated transconductance at
zero temperature is quantized to integer multiples of $4e^2/h$,
where $e$ is the electron charge and $h = 2\pi \hbar$ is Planck's
constant. For a sinusoidal current with frequency $\om$ applied
across one of the wires in the junction, we find that Shapiro
plateaus appear in the time-averaged voltage $\la V_1 \ra$ across
that wire for any rational fractional multiple (in contrast to only
integer multiples in junctions of two wires) of $2e \la V_1
\ra/(\hbar \om)$. We also use our formalism to study junctions of
two $p$- and one $s$-wave wires. We find that the corresponding
Andreev bound state energies depend on the spin of the Bogoliubov
quasiparticles; this produces a net magnetic moment in such
junctions. The time variation of these magnetic moments may be
controlled by an external voltage applied across the junction. We
discuss experiments which may test our theory.

\end{abstract}

\maketitle

\section{Introduction}
\label{sec:intro}

Josephson junctions have constituted a fascinating aspect of
superconducting systems from the very beginning; see, for example,
Ref.~\onlinecite{ketterson}. Such junctions exhibit a variety of
phenomena, such as the dc Josephson effect (a constant current flows
in the absence of any voltage biases between the different
superconductors), the ac Josephson effect (an alternating current
flows in the presence of constant voltage biases), and Shapiro steps
(these appear as plateaus in plots of the average voltage versus
average current when the currents are made to vary periodically).
The physics of such junctions is known to rely crucially on the
pairing symmetry of its constituent superconductors. For example, a
junction of two $p$-wave superconductors exhibits a fractional
Josephson effect \cite{kit1,kwon} which manifests itself in a
fractional Josephson frequency $\om_J=eV/\hbar$ or the absence of
odd-integer Shapiro steps. The latter property of such
junctions has been used experimentally for the detection of
Majorana modes \cite{expmaj}. Furthermore, a junction of two
superconducting wires of $s$- and $p$-wave symmetries is known to
generate a magnetic moment at the interface whose time variation can
be controlled by an external applied voltage \cite{kwon,sengupta}.
Multiple junctions with $s$-wave superconductors have been
studied using a scattering matrix formalism \cite{riwar,yoko,savi1,savi2},
and voltage-induced Shapiro steps have been studied in a junction of
three $s$-wave superconductors~\cite{cuevas}.
However, no such studies have been carried out for multi-terminal
junctions involving unconventional superconductors.

In recent years, topological phases of matter have also been
studied extensively ~\cite{hasan,qi}. These are usually
characterized by bulk band structures which have non-zero values of
some topological invariant, such as the Chern number in
two-dimensional systems. The value of the topological invariant
determines several properties of the system such as the number of
boundary modes and their contribution to electron transport.
Recently it has been shown that Josephson junctions of three or more
superconductors can exhibit interesting topological
properties~\cite{yoko,riwar,peng,erik,meyer,zazu,xie1,xie2} as
follows. First, there is a Berry curvature associated with the
Andreev bound-state wave functions; this curvature contributes to
the Josephson currents. Second, the current-voltage relation can, in
certain situations, involve a Chern number which is given by the
integral of the Berry curvature over a two-dimensional space of the
superconducting phases. A three-terminal Josephson interferometer
has been realized experimentally and some topological transitions
have been observed recently~\cite{strambini}. However, such studies
have not been extended to multi-wire junctions involving both $s$-
and $p$-wave superconductors.

In this work we develop a scattering matrix based approach which
allows us to address transport properties of multi-wire Josephson
junctions involving both $s$- and $p$-wave superconducting wires;
our work therefore constitutes a generalization of similar
multi-wire junctions involving only $s$-wave superconductors. The
plan of our paper and the key results obtained are as follows. In
Sec.~\ref{sec:model}, we introduce the model of $N$ superconducting
wires with either $s$- or $p$-wave symmetries meeting at a junction
which is characterized by a scattering matrix $S$; we obtain an
expression for the Andreev bound-state energies presented in terms
of $S$ and the pairing phases~\cite{been}. It is well known that for
$N \ge 3$, Berry curvature terms appear and can contribute to the
Josephson currents in the different wires; we show that for
junctions involving both $s$- and $p$- wave wires, such terms, along
with the Andreev bound-state energies, have a non-trivial dependence
on the spin of the quasiparticles. Second, we provide a
detailed study of the dc and ac Josephson effects in junctions with
all $s$- or all $p$-wave wires. We show that a constant voltage
$V_j$ applied across one of the wires (say the $j^{\rm th}$ wire) in
such a junction leads to a finite constant current in a different wire
(say the $i^{\rm th}$ wire) which receives contributions from the Berry
curvature terms and leads to a quantized zero temperature integrated
transconductance $G_{ij} = \int d\phi_i/(2\pi) (d \la I_i \ra
/dV_j) = 4e^2 p/h$, where $\phi_i$ is the superconducting phase of
the $i^{\rm th}$ wire, $p$ is an integer, and $\la .. \ra$
denotes time average over a time period $T= h/(2 e V_j)$. We also
consider an $RC$ circuit in which the Josephson current in each wire
flows in parallel with a resistance and a capacitance. We show that
if the applied current in one of the wires has a sinusoidally
varying term characterized by a frequency $\om$, Shapiro plateaus
can appear in the plot of the time-averaged voltage $\la V_1
\ra$ versus the average current in that wire. Our results show
that Shapiro plateaus for such junctions occur when $\la V_1
\ra$ and $\om$ satisfy $2 e \la V_1 \ra/(\hbar
\om)= m/l$, for integers $m$ and $l$. This indicates that plateaus
can appear when $2 e \la V_1 \ra/(\hbar \om)$ is any rational
fraction (in contrast to only integer values for standard Shapiro plateaus
in two-terminal junctions), and this may lead, in principle, to a
devil's staircase structure of such plateaus~\cite{maiti,shuk}. In
Sec.~\ref{sec:three}, we discuss the case of three $s$-wave
superconducting wires in detail; the Andreev bound states can be
analytically found in this case. We will take a simple example of an
$S$-matrix which is time-reversal symmetric and a randomly generated
example of an $S$-matrix which is not time-reversal symmetric; both
give rise to a Berry curvature but the Chern number (given by a
two-dimensional integral of the Berry curvature) is zero in the
first case and non-zero in the second case. We also discuss the
cases of three $p$-wave wires, and two $p$-wave and one $s$-wave
wire; in the latter case, the energies of spin-up and -down
Andreev bound states are not identical. In Sec.~\ref{sec:numerics},
we present numerical results for different three-wire systems. For
the case of three $s$-wave superconducting wires, we first discuss
the ac Josephson effect and find how this can clearly show the
effects of the Berry curvature. We then show that Shapiro plateaus
can appear in the plot of the voltage in a particular wire versus
the dc part of the current in the same wire when the current has an
ac part which varies sinusoidally with a frequency $\om$. We find
that Shapiro plateaus can appear at both integer and fractional
multiples of $\hbar \om /(2e)$. We provide an understanding of the
widths of the Shapiro plateaus by relating them
to the Fourier transforms of the energies of the Andreev
bound states. Similar Shapiro plateaus appear in the case of three
$p$-wave wires. For a system of two $p$-wave and one $s$-wave wires,
the asymmetry between the energies of spin-up and -down states
implies that there can be interesting spin-dependent effects; in
particular, we show that the junction region can have a net magnetic
moment whose time variation can be controlled by an external applied
voltage. We conclude in Sec.~\ref{sec:discussion} by summarizing our
results and suggesting some experimental tests of our results.

\section{Junction of $N$ superconducting wires}
\label{sec:model}

\subsection{Model}

\begin{figure}
\epsfig{figure=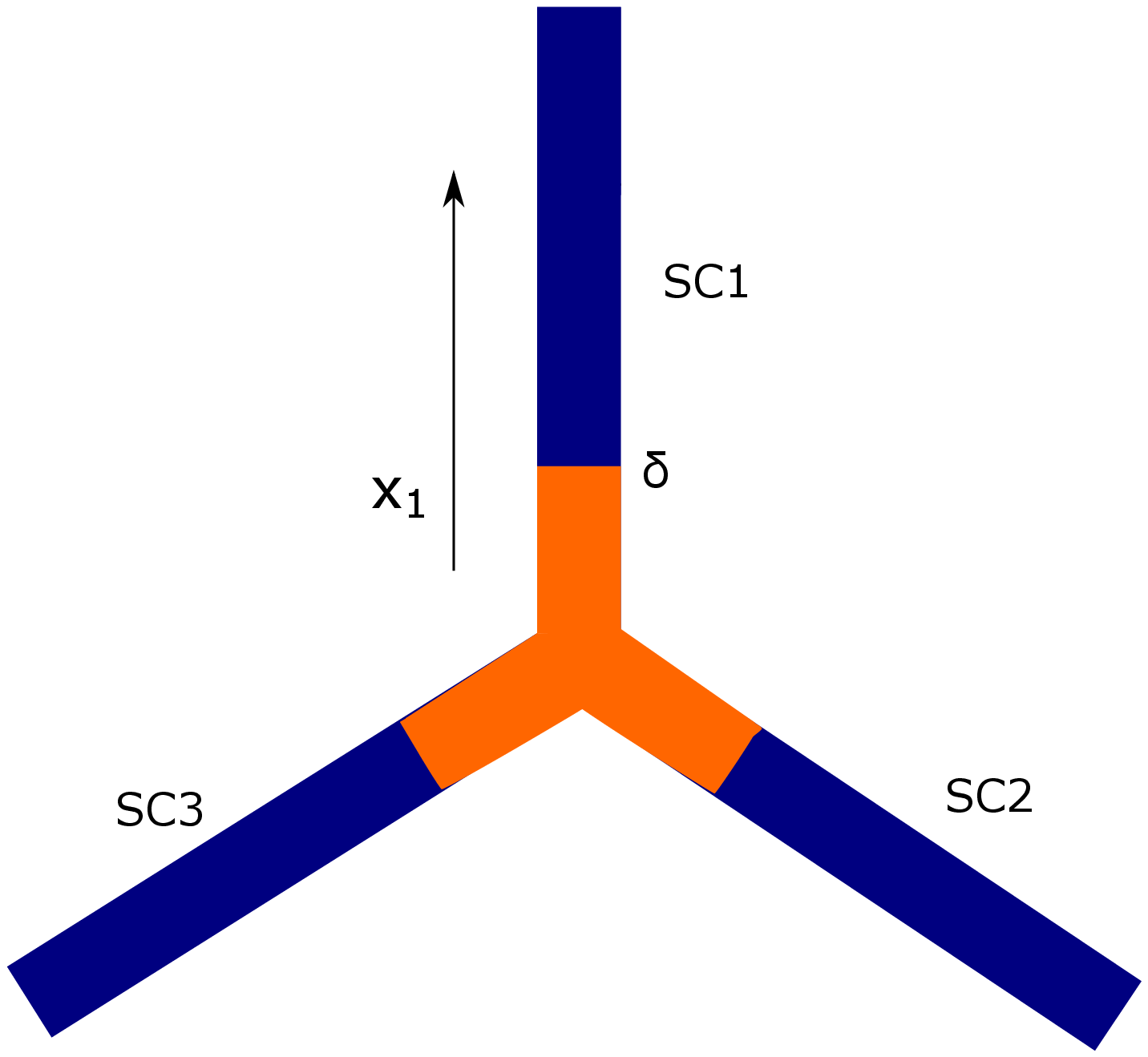,width=7cm}
\caption{Schematic picture of a junction of three superconducting wires.
The superconducting wires (marked SC1, SC2 and SC3) meet at a junction which
is a normal region which is characterized by an $S$-matrix. The coordinate
$x_1$ and the size of the junction $\de$ are indicated for the wire SC1.}
\label{fig:jj} \end{figure}

We consider a system consisting of $N$ wires which meet at a junction; a
schematic picture for $N=3$ is shown in Fig.~\ref{fig:jj}. Each wire,
labeled $i$, consists of a normal part (shown in orange color) where the
coordinate denoted by $x_i$ increases from zero (which is exactly at the
junction of the $N$ wires) to a small value $\de$. Beyond $x_i = \de$, the
wire is superconducting (shown in dark blue color). Beyond a large length, the
superconducting part of the wire is connected to a normal metal lead which is
at a potential $V_i$, and there is an incoming current on that lead given by
$I_i$; these leads are not shown in Fig.~\ref{fig:jj}. We will assume that
$\de$ (the length of the normal part of each wire) is small enough that we can
approximately set $e^{i k_F \de} = 1$, where $k_F$ is the
Fermi momentum (taken to be the same in all the wires).

We will be interested in both $s$- and $p$-wave superconductors. We
therefore recall the following facts about such SCs; see, for instance,
Ref.~\onlinecite{soori}. In terms of the spin Pauli matrices $\si^{x,y,z}$,
the pairing part of the Hamiltonian of an $s$-wave SC has the second-quantized
form
\bea H_{pair} &=& \int dx ~[\De e^{i\phi} \left( \begin{array}{cc}
\Psi^\dg_\ua & \Psi^\dg_\da \end{array} \right) i \si^y \left( \begin{array}{c}
\Psi^\dg_\ua \\
\Psi^\dg_\da \end{array} \right) + {\rm H.c.}] \non \\
&=& \int dx ~[\De e^{i\phi} (\Psi^\dg_\ua \Psi^\dg_\da ~-~ \Psi^\dg_\da
\Psi^\dg_\ua) + {\rm H.c.}]. \label{swavesc} \eea
Equation~\eqref{swavesc} describes Cooper pairs which are in the spin-singlet 
state $|\ua \da \ra - |\da \ua \ra$.
In a $p$-wave SC, the pairing part of the Hamiltonian is given by
\bea H_{pair} &=& \int dx ~[- \frac{i\De e^{i\phi} }{k_F} \left(
\begin{array}{cc}
\Psi^\dg_\ua & \Psi^\dg_\da \end{array} \right) {\vec d} \cdot {\vec \si} ~
i \si^y \frac{\pa}{\pa x} \left( \begin{array}{c} \Psi^\dg_\ua \\
\Psi^\dg_\da \end{array} \right) \non \\
&& ~~~~~~~~+ {\rm H.c.}], \label{pwavesc} \eea where ${\vec d}$ is a
unit vector; for ${\vec d} = {\hat x}, ~{\hat y}$ and ${\hat z}$,
the Cooper pairs are in the spin triplet states $|\ua \ua \ra - |\da
\da \ra$, $|\ua \ua \ra + |\da \da \ra$ and $|\ua \da \ra + |\da \ua
\ra$ respectively. (The factor of $-(i/k_F)\pa /\pa x$ in
Eq.~\eqref{pwavesc} gives $\pm 1$ if we consider electrons near the
Fermi momenta $\pm k_F$). We will restrict ourselves
to the case where all $p-$wave superconducting wires have the same
orientation of $\vec d$. In the absence of an external magnetic field,
we can then choose $\vec d \parallel {\hat z}$ without loss of generality.
In this case, Eq.~\eqref{pwavesc} takes the form \bea H_{pair} &=&
\int dx ~[-\frac{i\De e^{i\phi}}{k_F} (\Psi^\dg_\ua \frac{\pa}{\pa
x} \Psi^\dg_\da + \Psi^\dg_\da \frac{\pa}{\pa x} \Psi^\dg_\ua) \non \\
&& ~~~~~~~~+ {\rm H.c.}]. \label{pwavezsc} \eea Thus the Cooper
pairs are in the spin triplet state with $S^z = 0$; hence Cooper
pairs in $s$- and $p$-wave SC wires will have the same value of
$S^z$. This will make it possible to study a system with both
$s$-wave and $p$-wave SC wires since all the wires will be
compatible with each other at the junction where an electron or hole
from one of the wires can scatter into another wire. We note
that the choice of pairing in the $S^z=0$ channel for all the $p$-wave
SC wires in the junction does not lead to a loss of generality for the 
following reasons. First, for junctions with a single $p$-wave wire, 
the choice of pairing in the $S^z=0$ sector is merely a choice of the 
spin quantization axis to be along $\vec d \parallel {\hat z}$
\cite{leggettrmprev}; it does not alter the physical properties of the
junction. Second, the choice of the same direction $\vec d$ for the
different $p$-wave wires is not artificial. It is well known that
the direction of $\vec d$ depends on the material properties and
geometry of the wires; hence multiple wires constructed out of
the same material are expected to have $\vec d$ in the same direction.
Finally, a junction between two $p$-wave superconductors with
orthogonal $d$-vectors is known not to support a Josephson current
at zero temperature \cite{btkpaper}. Thus, taking theses issues into
consideration we choose all the $p$-wave SC wires in our
junction to have pairings in the $S^z=0$ channel.

We first consider a system of $N$ wires which are all $s$-wave SCs.
For spin-up quasiparticles, the annihilation operators are given by
superpositions of $\Psi_\ua$ and $\Psi^\dg_\da$, and we will denote
the corresponding wave functions by $\psi_{e\ua}$ and $\psi_{h\da}$.
For spin-down quasiparticles, the annihilation operators are
superpositions of $\Psi_\da$ and $\Psi^\dg_\ua$, and the
corresponding wave functions are $\psi_{e\da}$ and $\psi_{h\ua}$. We
introduce a symbol $\si = +1 ~(-1)$ to denote spin up (spin down)
respectively, and ${\bar \si} = -1 ~(+1)$ if $\si = +1 ~(-1)$.
Hence, the wave function of a spin-$\si$ quasiparticle is given by a
combination of $\psi_{e\si}$ and $\psi_{h\bar \si}$. For a spin-up
electron, let $\phi_i$ be the phase of the SC pairing amplitude
$\De_i$ on wire $i$, namely, $\De_i = \De e^{i\phi_i}$. We will take
the magnitude of the pairing, $|\De_i| = \De$, to be the same in all
the wires. Due to the spin-singlet nature of the Cooper pairs in an
$s$-wave SC, spin-down electrons will have the pairing amplitude
$\De_i = - \De e^{i\phi_i}$; this is clear from Eq.~\eqref{swavesc}.
We can therefore write the pairing amplitude as $\si \De
e^{i\phi_i}$ for spin-$\si$ electrons on wire $i$.

We now recall the derivation~\cite{been} of the energy $E_\si$ of an Andreev
bound state (ABS) which lies in the SC gap, i.e., $|E_\si| < \De$. To this end,
we assume that the normal region lying at the center of the system (i.e., the 
orange region in Fig.~\ref{fig:jj}) is characterized by a unitary $N \times N$
scattering matrix $S$. (If we impose time-reversal symmetry, $S$ will also be a
symmetric matrix). We will assume that $S$ does not depend on the energy; this
is because the magnitude of the pairing, $\De$, is typically much smaller than
the Fermi energy $E_F = \hbar^2 k_F^2/(2m)$, and we can therefore take $S$
to be constant in the range of energies from $- \De$ to $+ \De$ (note that
the energies are defined taking $E_F$ to be the zero level). The fact that
the central region is normal means that the form of $S$ should be the same
for spin-up and -down electrons even though electrons may have different
pairing amplitudes in the SC regions. Furthermore, the region being normal
implies that the incoming and outgoing electron wave functions in wire $i$,
$\psi_{e\si i}^{in} (E_\si)$ and $\psi_{e\si i}^{out} (E_\si)$, are related as
\beq \psi_{e\si i}^{out} (E_\si) ~=~ \sum_{j=1}^N S_{ij} \psi_{e\si j}^{in}
(E_\si), \label{eij} \eeq
the incoming and outgoing hole wave functions, $\psi_{h{\bar \si} i}^{in}
(E_\si)$ and $\psi_{h{\bar \si} i}^{out} (E_\si)$, are related as
\beq \psi_{h{\bar \si} i}^{out} (E_\si) ~=~ \sum_{j=1}^N S^*_{ij} \psi_{h{\bar
\si} j}^{in} (E_\si), \label{hij} \eeq
and electron and hole wave functions are not coupled to each other through $S$.
Next, when an outgoing electron (hole) in the normal region in wire $i$
strikes the junction with the SC at $x_i = \de$, it is Andreev reflected back
to the normal region as an incoming hole (electron). Namely~\cite{been},
\bea \psi_{e\si i}^{in} (E_\si) ~=~ a_\si (E_\si) \si e^{i \phi_i} \psi_{h{\bar
\si}
i}^{out} (E_\si), \non \\
\psi_{h{\bar \si}i}^{in} (E_\si) ~=~ a_\si (E_\si) \si e^{-i \phi_i}
\psi_{e\si i}^{out} (E_\si), \label{swave} \eea where \beq a_\si
(E_\si) ~=~ \frac{E_\si ~-~ i \sqrt{\De^2 - E_\si^2}}{\De}. \label{ae} \eeq
(Note that we are ignoring any phases picked up by
the electrons or holes while propagating between the junction at $x_i = 0$
and the SC at $x_i = \de$ due to the approximation $e^{ik_F \de} = 1$).
In Eq.~\eqref{ae} we note that the real part of $a_\si (E_\si)$ can
be positive, negative or zero, while the imaginary part can only be
negative or zero; this is important since the eigenvalue equations
given below will only fix the value of $a_\si^2 (E_\si)$, and we
then have to take the appropriate square root of that to obtain
$a_\si (E_\si)$. Combining Eqs.~(\ref{eij}-\ref{swave}), we find
that \beq \psi_{e\si i}^{out} (E_\si) ~=~ a_\si^2 (E_\si)
~\sum_{j,k=1}^N ~S_{ij} e^{i\phi_j} S^*_{jk} e^{-i \phi_k}
~\psi_{e\si k}^{out} (E_\si). \label{abs1} \eeq Introducing an
$N$-dimensional column $\psi_\si (E_\si)$ whose entries are given by
$\psi_{e \si i}^{out} (E_\si)$, a diagonal matrix $e^{i\phi}$ whose
diagonal entries are given by $e^{i\phi_i}$, and its inverse matrix
$e^{-i\phi}$, Eq.~\eqref{abs1} takes the form of an eigenvalue
equation \beq S e^{i\phi} S^* e^{-i\phi} \psi_\si (E_\si) ~=~
\frac{1}{a_\si^2 (E_\si)}~ \psi_\si (E_\si). \label{abs2} \eeq It is
clear from Eq.~\eqref{abs2} that the ABS energies and the
corresponding wave functions do not change if any of the phases
$\phi_i$ are shifted by $2\pi$. In addition, the ABS energies and
wave functions remain unchanged if all the phases $\phi_i$ are
shifted by the same constant, since that constant will cancel out
between $e^{i \phi}$ and $e^{-i \phi}$. As a result, we have the
identities
\bea \sum_{i=1}^N ~\frac{\pa E_\si}{\pa \phi_i} &=& 0, \label{sume} \\
\sum_{i=1}^N \frac{\pa \psi_\si}{\pa \phi_i} &=& 0, ~~~{\rm and}~~~
\sum_{i=1}^N \frac{\pa \psi_\si^\dg}{\pa \phi_i} = 0. \label{sumpsi} \eea
We will use this symmetry to set one of the phases (for example, $\phi_3$
for a three-wire system) equal to zero in many of the calculations.

Equation~\eqref{abs2} implies that
\beq S e^{i\phi} S^* e^{-i\phi} S e^{i\phi} \psi_\si^* (E_\si) ~=~ \frac{1}{
a_\si^{*2} (E)} S e^{i\phi} \psi_\si^* (E_\si). \label{absx1} \eeq
Since $a_\si^{*2} (E_\si) = [(-E_\si - i \sqrt{\De^2 - E_\si^2})/\De]^2$,
Eqs.~\eqref{abs2} and \eqref{absx1} imply that if there is an ABS at energy
$E_\si$ with wave function $\psi_\si (E_\si)$, there must be an ABS at energy
$-E_\si$ with wave function $\psi_\si (-E_\si) = S e^{i\phi} \psi_\si^*
(E_\si)$. An exception to this statement can occur if $a_\si^2 (E_\si)$ is
real in which case there may be only one ABS with no degeneracy. In particular,
if $a_\si^2 (E_\si) = -1$, there may be only one state lying at $E_\si = 0$,
and if $a_\si^2 (E_\si)
= 1$, there may be only one state lying at $E_\si^2 = \De^2$ (however, this
should not really be considered to be a bound state since its energy lies at
the edge between the SC gap and the bulk states, and its localization length
is therefore large).

We note that Eq.~\eqref{abs2} has the same form for $\si = \pm 1$.
Namely, the ABS energies $E_\si$ and wave functions $\psi_\si$ are
identical for spin-up and -down quasiparticles in a system in
which all the wires are $s$-wave SCs. Hence each energy will have a
two-fold degeneracy.

Next, we discuss a system in which some of the SC wires are $s$-wave
and the others are $p$-wave~\cite{kwon,sengupta}. Comparing
Eqs.~\eqref{swavesc} and \eqref{pwavezsc}, we see that the pairing
amplitudes for spin-up and -down electrons have opposite signs
for an $s$-wave SC but the same sign for a $p$-wave SC. Hence if
wire $i$ is an $s$-wave SC, Eq.~\eqref{swave} holds. But if wire $i$
is a $p$-wave SC, we find that the factors of $\si$ do not appear;
however one of the equations in Eq.~\eqref{swave} picks up a minus
sign. Namely, we find that \bea \psi_{e\si i}^{in} (E_\si) ~=~ -
~a_\si (E) e^{i \phi_i} ~\psi_{h{\bar \si} i}^{out} (E_\si), \non \\
\psi_{h{\bar \si} i}^{in} (E_\si) ~=~ a_\si (E_\si) e^{-i \phi_i}
~\psi_{e\si i}^{out} (E_\si). \label{pwave} \eea This leads us to
define a diagonal matrix $\eta$ whose $i$-th diagonal entry $\eta_{ii}$
is equal to $+1$ ($-1$) if the $i$-th SC wire is $s$-wave ($p$-wave).
To go from spin-up quasiparticles to spin-down
quasiparticles, we have to change $e^{i\phi_i} \to - \eta_{ii} e^{i\phi_i}$
and $e^{-i\phi_i} \to - \eta_{ii} e^{-i\phi_i}$ on wire $i$. We thus see that
to account for spin, it is useful to consider a diagonal matrix which is
equal to $-\eta$. We then find that the ABS
energy is given by the eigenvalue equation \beq S \eta e^{i \phi}
S^* e^{-i\phi} \psi_\si (E_\si) ~=~ \frac{1}{a_\si^2 (E_\si)}
\psi_\si (E_\si), \label{abs4u} \eeq
for spin-up quasiparticles ($\si = +1$), and
\beq S e^{i \phi} S^* \eta e^{-i\phi} \psi_\si
(E_\si) ~=~ \frac{1}{a_\si^2 (E_\si)} \psi_\si (E_\si), \label{abs4d} \eeq
for spin-down quasiparticles ($\si = -1$). We
then see that for each value of $\si$, the ABS energies have an $E
\to - E$ symmetry only if all the SC wires are $s$-wave or all are
$p$-wave. If some of them are $s$-wave and some are $p$-wave, there
is no $E \to - E$ symmetry in general.

We now note that Eq.~\eqref{abs4u} implies \beq S e^{i\phi} S^* \eta
e^{-i\phi} S e^{i\phi} \psi_\si^* (E_\si) ~=~ \frac{1}{a_\si^{*2}
(E_\si)} S e^{i\phi} \psi_\si^* (E_\si). \label{absx2} \eeq Hence,
if there is an ABS at energy $E$ with wave function $\psi_\si (E)$
for spin-up quasiparticles, there must be an ABS at energy $-E$ with
wave function $\psi_{\bar \si} (-E) = S e^{i\phi} \psi_\si^* (E)$
for spin-down quasiparticles. Thus the combined energy spectrum for
spin-up and -down quasiparticles will always have an $E \to - E$
symmetry, even if the spectra for spin-up or spin-down
quasiparticles separately do not have such a symmetry.

\subsection{Two-wire system}

In this subsection, we consider a two-wire system to show
that our formalism reproduces the known result for both conventional
(two $s$-wave wires) and unconventional (one $s$- and one $p$-wave
wire or two $p$-wave wires) junctions. To this end, we first note
that the unitarity of the $2 \times 2$ scattering matrix implies that
$|S_{11}|=|S_{22}|$ and $|S_{12}|=|S_{21}|$. We then find the
following results for the two ABS energies. For two $s$-wave SC
wires, the ABS energies are given by \beq E_\si ~=~ \pm \De ~\sqrt{1
~-~ |S_{12}|^2 \sin^2 (\frac{\phi_1 - \phi_2}{2})} \label{abs9} \eeq
for both spin-up and -down quasiparticles. For two $p$-wave
wires, the ABS energies are \beq E_\si ~=~ \pm \De ~|S_{12}| \sin
(\frac{\phi_1 - \phi_2}{2}) \label{abs10} \eeq for both spin-up and
spin-down quasiparticles. If one wire is an $s$-wave SC and the
other is $p$-wave, the ABS energies are
\bea E_\si &=& \si ~{\rm sgn} \left( \sin (\phi_1 - \phi_2) \right) \non \\
&& \times ~\De ~\sqrt{\frac{1 \pm \sqrt{1 ~-~ |S_{12}|^4 \sin^2 (\phi_1
- \phi_2)}}{2}}, \label{abs11} \eea
where $\rm sgn$ denotes the signum function.

To compare the expressions in Eqs.~(\ref{abs9}-\ref{abs11}) with those
given in Ref.~\onlinecite{kwon}, we have to note the following. At a point $x$
in a SC wire, an $s$-wave SC has a pairing of the form given in
Eq.~\eqref{swavesc} while a $p$-wave SC has a pairing of the form given in
Eq.~\eqref{pwavesc}. On comparing the two expressions, we see that an $s$-wave
pairing phase $\phi$ is insensitive to the sign of the coordinate $x$ while a
$p$-wave pairing phase depends on the sign of $x$ because of the $\pa /\pa x$.
As mentioned at the beginning of this section, we are taking the
coordinates $x_i$ in every wire to increase from zero at the junction; this is
a convenient choice for three or more wires. However, for a two-wire system, it
is conventional to take $x$ to go from $-\infty$ to $\infty$ with the junction
being at $x=0$; hence the coordinate increases from zero to $\infty$ in one of
the wires and decreases from zero to $-\infty$ in the other wire. We can
change our notation to agree with this convention by changing the coordinate
in, say, wire 1 from $x_1 \to - x_1$. If wire 1 has $p$-wave pairing,
this must be compensated by changing the phase $\phi_1 \to \phi_1 + \pi$.
No such change in the phase is required if wire 1 has $s$-wave pairing.
Hence, for two $s$-wave SC wires the ABS energies are still given by
Eq.~\eqref{abs9}. But for two $p$-wave wires, we have to shift either $\phi_1$
or $\phi_2$ by $\pi$. This changes the expression for the ABS energies from
Eq.~\eqref{abs10} to
\beq E_\si ~=~ \pm \De ~|S_{12}| \cos (\frac{\phi_1 - \phi_2}{2}).
\label{abs12} \eeq
If wire 1 is $s$-wave and wire 2 is $p$-wave, we do not have to change $\phi_1$
and $\phi_2$, and the ABS energies are again given by Eq.~\eqref{abs11}.
But if wire 1 is $p$-wave and wire 2 is $s$-wave, we have to shift $\phi_1$
by $\pi$ and the ABS energies are then given by Eq.~\eqref{abs11} multiplied by
$-1$. The expressions in Eqs.~\eqref{abs9}, \eqref{abs12} and \eqref{abs11}
(up to a sign) agree with those given in Ref.~\onlinecite{kwon}.

\subsection{Berry curvature}

Returning to the case of $N$ wires, we now look at the wave function
$\psi_{n,\si} (E_{n,\si}) = \psi_{e,n,\si}^{out} (E_{n,\si})$ given
by Eq.~\eqref{abs2} for an ABS with spin $\si$ and energy
$E_{n,\si}$ in the band labeled as $n$ (where $n=1,2,\cdots,N$).
Following Ref.~\onlinecite{riwar}, we define the Berry curvature
matrix \beq B_{n,\si,ij} (\phi_1, \phi_2, \cdots, \phi_N)~=~ - 2
~{\rm Im} \left[ \frac{\pa \psi_{n,\si}^\dg}{\pa \phi_i} \frac{\pa
\psi_{n,\si}}{\pa \phi_j} \right]. \label{berry} \eeq This is a real
antisymmetric matrix with the following properties.
Equation~\eqref{sumpsi} implies that each row and each column of
$B_{n,\si,ij}$ adds up to zero, i.e., \beq \sum_{i=1}^N
~B_{n,\si,ij} ~=~ \sum_{j=1}^N ~B_{n,\si,ij} ~=~ 0. \label{bij} \eeq
These identities along with the antisymmetry imply that, for each
value of $n$ and $\si$, the matrix $B_{n,\si,ij}$ contains only
$(N-1)(N-2)/2$ independent real parameters which we can take to be
the values of $B_{n,\si,ij}$ for $1 \le i < j \le N-1$. Thus the
Berry curvature can be non-zero only if $N \ge 3$. We note here,
that in contrast to junctions of all $s$-wave wires or all $p$-wave
wires, the Berry curvature for a mixed $s-p$ junction depends on
$\si$. In this paper, we will use the technique discussed in
Ref.~\onlinecite{fukui} to calculate the Berry curvature. [Note that
Eq.~\eqref{berry} would have given the same values of $B_{n,\si,ij}$
if we had used $\psi_{e,n,\si}^{in}$ instead of $\psi_{n,\si}
(E_{n,\si}) = \psi_{e,n,\si}^{out}$, since these are related by the
matrix $S$ as in Eq.~\eqref{eij} and $S$ is independent of the
$\phi_i$'s].

Holding the phases $\phi_3, \phi_4, \cdots, \phi_N$ fixed, we can define a
Chern number $Ch_{n,\si,12}$ by integrating $B_{n,\si,12}$ over $\phi_1$ and
$\phi_2$,
\beq Ch_{n,\si,12} ~=~ \frac{1}{2\pi} ~\int_0^{2\pi} \int_0^{2\pi} d\phi_1
d\phi_2~ B_{n,\si,12}. \label{c12} \eeq
This is always quantized to have integer values.

Following Eq.~\eqref{absx2}, we can derive a relation between the
Berry curvature for a positive energy, spin-up band (denoted as
$n,~\si=+1$), and a negative energy, spin-down band (denoted as
$n',~\si=-1$). The wave functions in the two bands are related as
$\psi_{n',-1} (-E) = S e^{i\phi} \psi^*_{n,+1} (E)$, where $E > 0$.
Let us denote the $j$-th component of the wave function $\psi_{n,+1}
(E)$ by $\psi_{n,j}$, where $j=1,2,\cdots,N$. We can then show that
\bea B_{n',-1,ij} (-E) &=& - ~B_{n,+1,ij} (E) \non \\
&& -~ \frac{\pa |\psi_{n,j}|^2}{\pa \phi_i} ~+~ \frac{\pa |\psi_{n,i}|^2}{\pa
\phi_j}. \label{bsym} \eea
[We will see later that the last two terms in Eq.~\eqref{bsym} are sometimes
much smaller than the first term; we then have $B_{n',-1,ij} (-E) \simeq -
B_{n,+1,ij} (E)$]. Equation~\eqref{bsym} implies the exact relation
\beq Ch_{n',-1,ij} ~=~ -~ Ch_{n,+1,ij}. \label{csym} \eeq

For any value of the pairing phases $\phi_i$, we can derive a sum
rule for the ABS energies and Berry curvatures of all the bands and
the two possible spins. According to Eq.~\eqref{abs4u}, $\psi_{n,j}$
is the $j$-th component of the $n$-th eigenstate of the unitary
matrix $U = S \eta e^{i\phi} S^* e^{-i\phi}$. The orthonormality of
the eigenstates of a unitary matrix implies that $\sum_n
|\psi_{n,j}|^2 = 1$ for each value of $j$; hence $\sum_n \pa
|\psi_{n,j}|^2 /\pa \phi_i = 0$ for each value of $i$ and $j$.
Equation~\eqref{bsym} then leads to the identity
\beq \sum_{\si = \pm 1} \sum_{n=1}^N B_{n,\si,ij} ~=~ 0. \label{bsum} \eeq
Furthermore, since the energies of the $\si=+1$ bands have opposite signs to
the energies of the $\si=-1$ bands (Eq.~\eqref{absx2}), we have \beq
\sum_{\si = \pm 1} \sum_{n=1}^N E_{n,\si} ~=~ 0. \label{esum} \eeq

We now consider a situation in which a voltage $V_i$ is applied to
wire $i$; the phase $\phi_i$ then varies in time according
to~\cite{ketterson} \beq {\dot \phi}_i ~=~ \frac{2e}{\hbar} ~V_i,
\label{phidot} \eeq where $\dot \phi_i \equiv d\phi/dt$. Next, the
contribution of a particular ABS with energy $E_{n,\si}$ to the
Josephson current in the $i$-th wire is given by~\cite{riwar}
\bea I_i &=& \frac{1}{2} ~\sum_{\si=\pm 1} \sum_{n=1}^N ~[f(E_{n,\si}) ~-~
\frac{1}{2}] \non \\
&& \times~ \left[\frac{2e}{\hbar} ~\frac{\pa E_{n,\si}}{\pa \phi_i}
~-~ 2 e ~ \sum_{j=1}^N B_{n,\si,ij} {\dot \phi}_j \right].
\label{ii1} \eea
(The prefactor of $1/2$ has been included to avoid double counting of
spin; see Ref.~\onlinecite{kwon}).
Note that while summing over $n$ and $\si$, we have
included the effect of a temperature $T$ through the Fermi function
\beq f(E_{n,\si}) ~=~ \frac{1}{e^{\beta E_{n,\si}} +1}, \eeq where
$\beta = 1/(k_B T)$, $k_B$ is the Boltzmann constant, and we have
taken the chemical potential to lie at the center of the
SC gap; this will be discussed in detail in Sec.~\ref{sec:three}.
We can see that Eq.~\eqref{ii1} satisfies current conservation,
$\sum_{i=1}^N I_i = 0$ due to Eqs.~\eqref{sume} and \eqref{bij}. In
Eq.~\eqref{ii1} we have introduced the temperature dependence as
$f(E_{n,\si}) - 1/2$ following Ref.~\onlinecite{riwar}. However,
Eqs.~(\ref{bsum}-\ref{esum}) imply that the value of $I_i$ would not
change if we dropped the factor of $1/2$.

\subsection{Josephson effects}

In this section, we study the ac Josephson effect in these
junctions. To this end, we first note that if $V_i = 0$, the
$\phi_i$'s are constant in time and we get constant currents given
by \beq I_i ~=~ \frac{1}{2} ~\sum_{n,\si} ~[f(E_{n,\si}) ~-~ \frac{1}{2}]
\left[\frac{2e}{\hbar} ~\frac{\pa E_{n,\si}}{\pa \phi_i}\right]. \eeq
This is called the dc Josephson effect. The Berry curvature
does not contribute in this case.

To study the ac Josephson effect, we first take the $V_i$ to be
time-independent non-zero constants leading to $\phi_i = (2e/\hbar) V_i t$ and
\bea I_i &=& \frac{1}{2} ~\sum_{n,\si} ~[f(E_{n,\si}) ~-~ \frac{1}{2}] \non \\
&& \times ~\left[\frac{2e}{\hbar} ~\frac{\pa E_{n,\si}}{\pa \phi_i} ~-~ \frac{4
e^2}{\hbar} ~\sum_{j=1}^N B_{n,\si,ij} V_j \right]. \label{curr2} \eea
These currents vary with time since $\phi_i$ and therefore $B_{n,\si,ij}$ vary 
with time. Equation~\eqref{curr2} can be written in units of energy (eV) as
\bea \frac{\hbar I_i}{2e} &=& \frac{1}{2} ~\sum_{n,\si} ~[f(E_{n,\si}) ~-~
\frac{1}{2}] \non \\
&& \times ~\left[\frac{\pa E_{n,\si}}{\pa \phi_i} ~-~\sum_{j=1}^N B_{n,\si,ij}~
2e V_j \right]. \label{curr3} \eea

Let us consider the case where only $V_1 \ne 0$ and $V_2 = V_3 =
\cdots = 0$. Then $\phi_1 = (2e/\hbar) V_1 t$ and $\phi_2, \phi_3,
\cdots$ are some constants. Since the system remains invariant when
$\phi_1 \to \phi_1 + 2 \pi$ keeping the other $\phi_i$'s fixed, we
see that $E_{n,\si}$ and $B_{n,\si,ij}$ vary in time with a period
$T = 2\pi \hbar/(2 e V_1)$. Equation~\eqref{curr2} implies that the total
charge flowing in wire $i$ in one time period $T$ is given by
\bea && Q_i ~=~ \frac{1}{2} ~\sum_{n,\si} ~[f(E_{n,\si}) ~-~ \frac{1}{2}]
\non \\
&& \times \left[ \frac{2e}{\hbar} \int_0^T dt ~\frac{\pa E_{n,\si}}{\pa \phi_i}
- \frac{4 e^2 V_1}{\hbar} \int_0^T dt ~B_{n,\si,i1} \right]. \label{qt} \eea
The average current flowing over time $T$
is then given by \beq \la I_i \ra ~=~ \frac{Q_i}{T}. \label{qit1}
\eeq (We are interested in $\la I_i \ra$ since it gives the dc part
of the current). Since $E_{n,\si}$ is a periodic function of
$\phi_1$, and $\phi_1$ varies linearly in time with a period $T$, we
see that \beq \int_0^T dt ~\frac{\pa E_{n,\si}}{\pa \phi_1} ~=~ 0.
\label{int1} \eeq This equation, along with $B_{n,\si,11} = 0$,
implies that $Q_1 = 0$; hence $\la I_1 \ra = 0$.

For the case of two wires ($N=2$), the facts that $E_{n,\si}$ is a
function of $\phi_1 - \phi_2$ and $B_{n,\si,ij} = 0$ can be used to
analytically show that $\la I_1 \ra = \la I_2 \ra = 0$. But for $N
\ge 3$, we find numerically that $\la I_i \ra$ is generally not
equal to zero for $i=2,3,\cdots,N$. Thus the application of a
constant voltage bias on one wire produces a constant current in all
the other wires (in addition to a time-dependent current which gives
zero when integrated over time $T$). This phenomenon of
transconductance (defined as $G_{ij} = \la I_i \ra/V_j$ where $i \ne
j$) has been studied earlier~\cite{erik,meyer,xie1,xie2}. However,
those papers discussed this phenomenon when incommensurate voltage
biases are applied to two of the wires (the time-averaged
transconductance is then quantized), whereas we have shown here that
the transconductance is non-zero even if a voltage bias is applied
to only one wire. The transconductance for a particular set of
values of $\phi_i$ is not quantized in our case; however, the
integral of the transconductance over one of the phases $\phi_i$ is
quantized at zero temperature as we will now show.

If we hold $\phi_3, \phi_4, ...$ fixed, the invariance of
$E_{n,\si}$ under $\phi_2 \to \phi_2 + 2\pi$ implies that
$\int_0^{2\pi} d\phi_2 \pa E_{n,\si}/ \pa \phi_2 = 0$, while the
linear variation of $\phi_1$ with $t$ implies that $(1/T) \int_0^T
dt B_{n,\si,12} = (1/2\pi) \int_0^{2\pi} d\phi_1 B_{n,\si,12}$.
Using Eqs.~\eqref{c12}, \eqref{qt} and \eqref{qit1}, we obtain, at
zero temperature where $f(E_{n,\si}) -1/2 = - {\rm sgn} (E_{n,\si})
/2$, \bea && \int_0^{2\pi} ~\frac{d\phi_2}{2\pi} ~\frac{\la I_2
\ra}{V_1} ~=~ \frac{2 e^2}{h} ~\sum_{n,\si} ~\left(-~ \frac{{\rm
sgn} (E_{n,\si})}{2}
\right) ~Ch_{n,\si,12} \non \\
&& =~ \frac{4 e^2}{h} ~\sum_n ~\left(-~ \frac{{\rm sgn}
(E_{n,+1})}{2} \right) ~Ch_{n,+1,12}, \label{i2c12} \eea where $h =
2\pi \hbar$, and we have used Eq.~\eqref{csym} and the fact that
$E_{n',-1} = - E_{n,+1}$ to derive the second line in Eq.~\eqref{i2c12}
from the first line (i.e., we have removed the sum over $\si$ and
changed the prefactor from 2 to 4). Since $Ch_{n,+1,12}$ is an
integer for all values of $n$, and Chern numbers of positive and
negative energy bands have opposite signs, we see that the integral
of the transconductance over $\phi_2$ is quantized in units of
$4e^2/h$.

\begin{figure}
\epsfig{figure=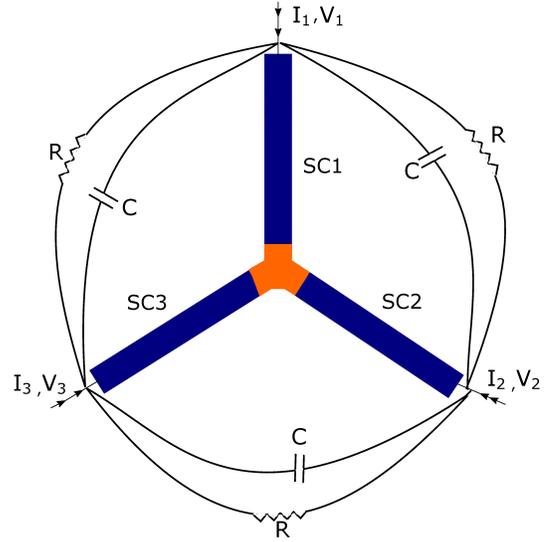,width=7cm}
\caption{Schematic picture of an $RC$ circuit with three superconducting
wires (marked SC1, SC2 and SC3) which meet at a junction with an $S$-matrix.
The superconducting wires are in parallel to resistances marked $R$ and
capacitances marked $C$. The voltage biases $V_i$ and incoming currents
$I_i$ are shown.} \label{fig:rc} \end{figure}

\subsection{Shapiro plateaus}

In this section, we consider a sinusoidal applied voltage,
which, for two-wire junctions, is well known to lead to Shapiro
plateaus~\cite{ketterson,maiti,shuk}. To understand this, it is
common to consider an $RC$ circuit in which a resistance $R_{ij}$
and a capacitance $C_{ij}$ are placed between every pair of leads
$i$ and $j$ so that the currents flowing through them are in
parallel to the Josephson currents; a schematic picture of this for
a three-wire system is shown in Fig.~\ref{fig:rc}. The equations for
the current in the $i$-th wire are then modified from Eq.~\eqref{ii1} to
\bea I_i &=& \frac{1}{2} \sum_{n,\si} ~[f (E_{n,\si}) - \frac{1}{2}] \left[
\frac{2e}{\hbar} \frac{\pa E_{n,\si}}{\pa \phi_i} - 2 e \sum_{j \ne i}
B_{n,\si,ij} {\dot \phi}_j \right] \non \\
&& + ~\sum_{j\ne i} ~[ C_{ij} ({\dot V}_i - {\dot V}_j) ~+~
\frac{V_i - V_j}{R_{ij}} ], \label{ii2} \eea
where we have used the fact that $B_{n,\si,ij} = 0$ if $i = j$. For
convenience, let us assume that $R_{ij}=R$ and $C_{ij}=C$ for all pairs of
leads. Using Eq.~\eqref{phidot}, we then obtain \bea I_i &=& \frac{1}{2}
\sum_{n,\si} ~[f (E_{n,\si}) - \frac{1}{2}] \left[ \frac{2e}{\hbar} \frac{\pa
E_{n,\si}}{\pa \phi_i} - 2 e \sum_{j \ne i} B_{n,\si,ij} {\dot
\phi}_j \right] \non \\
&& + ~\frac{\hbar}{2e} ~\sum_{j\ne i} ~[C ({\ddot \phi}_i- {\ddot \phi}_j)
~+~ \frac{{\dot \phi}_i - {\dot \phi}_j}{R} ]. \label{ii3} \eea

We now consider the case where \beq I_1 ~=~ I ~+~ A \sin (\om t),
\label{i1} \eeq and $V_2 = V_3 = \cdots = 0$. Then $\phi_2, \phi_3,
\cdots$ are all constant in time, while $\phi_1$ will vary with
time. Equation~\eqref{ii3} then gives 
\bea && \frac{1}{2}~ \sum_{n,\si} ~[f (E_{n,\si}) - \frac{1}{2}] 
\left[\frac{2e}{\hbar} ~\frac{\pa E_{n,\si}}{\pa \phi_1} \right] \non \\
&& +~ \frac{(N-1)\hbar}{2e} ~[ C {\ddot \phi}_1 ~+~ \frac{{\dot \phi}_1}{R}]
~=~ I + A \sin (\om t), \label{ii4} \eea
where $N$ is the number of wires, and
\bea && \frac{1}{2} ~\sum_{n,\si} ~[f (E_{n,\si}) - \frac{1}{2}] \left[
\frac{2e}{\hbar} ~\frac{\pa E_{n,\si}}{\pa \phi_j} ~-~ 2 e B_{n,\si,j1}
{\dot \phi}_1 \right] \non \\
&& -~ \frac{\hbar}{2e} ~[C {\ddot \phi}_1 ~+~ \frac{{\dot \phi}_1}{R}] ~=~
I_j, \label{ii5} \eea
for $j \ne 1$. We observe that Eq.~\eqref{ii4} does not contain any Berry
curvature terms; this is therefore the simplest equation to solve. Given
some initial values of $\phi_1$ and ${\dot \phi}_1$ at time $t=0$, we can
solve this numerically. We can write Eq.~\eqref{ii4} in the form~\cite{maiti}
\bea && {\ddot \phi}_1 + \ga {\dot \phi}_1 + \frac{2e^2}{(N-1)\hbar^2 C}
\sum_{n,\si} ~[f (E_{n,\si}) - \frac{1}{2}] \left[ \frac{\pa E_{n,\si}}{\pa
\phi_1} \right] \non \\
&& = \frac{2e}{(N-1)\hbar C} ~[I ~+~ A \sin (\om t)], \label{ii6} \eea
where $\ga = 1/(RC)$ is a positive quantity.

We will now present a perturbative argument to understand how Shapiro
plateaus can arise from Eq.~\eqref{ii6}. (A similar procedure has been 
presented in Ref.~\onlinecite{likh} and applied in Ref.~\onlinecite{maiti}).
The perturbation 
parameter will be taken to be the SC pairing amplitude $\De$; note that the 
ABS energies $E_{n,\si} = 0$ if $\De = 0$, as we can see from Eq.~\eqref{ae}. 
To zeroth order in $\De$, therefore, the third term on the left hand side of
Eq.~\eqref{ii6} (namely, the Josephson current) is equal to zero. Note that
this term is a non-linear function of $\phi_1$; omitting this term therefore
gives a simple linear equation. After a long time, when a transient term
decaying as $e^{-\ga t}$ has gone to zero, the general solution of this
equation is given by
\bea \phi_1 &=& \al t ~+~ a \sin (\om t + \chi) ~+~ \phi_0, \non \\
{\rm where} ~~~~\al &=& \frac{2eI}{(N-1) \hbar \ga C}, \non \\
a &=& - ~\frac{2eA}{(N-1)\hbar C \om \sqrt{\ga^2 + \om^2}}, \non \\
\chi &=& \tan^{-1} (\frac{\ga}{\om}), \label{ii7} \eea
and $\phi_0$ is an arbitrary constant.

We can now use the above result to put back the third term on the left hand
side of Eq.~\eqref{ii6}. We do this as follows. For a fixed set values of
$\phi_2, \phi_3, \cdots$, we know that $E_{n,\si}$ is a periodic function of
$\phi_1$ with a period $2\pi$. We can therefore write
\bea \frac{\pa E_{n,\si}}{\pa \phi_1} &=& \sum_{l=-\infty}^\infty ~c_l ~e^{il
\phi_1}, \non \\
{\rm where}~~~~ c_l &=& \int_0^{2\pi} \frac{d\phi_1}{2\pi}~
\frac{\pa E_{n,\si}}{\pa \phi_1} ~e^{-il \phi_1}. \label{fourier} \eea
The Fourier coefficients $c_l$ are functions of $\phi_2, \phi_3, \cdots$. (The
coefficient $c_0$ must be equal to zero since we know that $\int_0^{2\pi}
d\phi_1 \pa E_{n,\si}/\pa \phi_1 = 0$. Also, since $\pa E_{n,\si}/\pa \phi_1$
is real, we must have $c_{-l}= c_l^*$; this implies that $|c_{-l}| = |c_l|$ for
all values of $l$). Substituting Eq.~\eqref{ii7} in Eq.~\eqref{fourier} and
using the identity~\cite{abram}
\beq e^{iz \sin \ta} ~=~ \sum_{m=-\infty}^\infty ~J_m (z) ~e^{im\ta}, \eeq
where $J_m (z)$ denotes the Bessel function of order $m$, we obtain
\beq e^{il\phi_1} ~=~ e^{il (\al t + \phi_0)} ~\sum_{m=-\infty}^\infty ~J_m
(la) ~e^{im(\om t + \chi)}. \label{bessel} \eeq
For a function $f(t)$, we define the long-time average value as
\beq \la f \ra ~=~ \lim_{T \to \infty} ~\frac{1}{T} ~\int_0^T ~dt ~f(t).
\label{avf} \eeq
Using the result
\bea \lim_{T \to \infty} ~\frac{1}{T} ~\int_0^T ~dt ~e^{i\ep t} &=& 0
~~~~{\rm if}~~~~ \ep \ne 0, \non \\
&=& 1 ~~~~{\rm if}~~~~ \ep = 0, \eea
we find from Eqs.~(\ref{fourier}-\ref{bessel}) that
\beq \la \frac{\pa E_{n,\si}}{\pa \phi_1} \ra ~=~ \sum_{l,m} ~c_l J_m (la)
e^{im\chi + i l \phi_0}, \label{eav} \eeq
whenever $\al = -(m/l) \om$ for a pair of integers $m$ and $l$, and is equal
to zero otherwise. On the other hand, Eqs.~\eqref{phidot} and \eqref{ii7} give
$(2e/\hbar) \la V_1 \ra = \al$. We therefore conclude that Eq.~\eqref{eav}
is non-zero only if $\la V_1 \ra$ is a rational multiple of $\hbar \om /(2e)$.

Next, if Eq.~\eqref{eav} is non-zero, we see from Eqs.~\eqref{i1} and
\eqref{ii6} that this effectively changes the dc part of $I_1$ from $I$ to
$I + (e/\hbar) \sum_{n,\si} [f (E_{n,\si})-1/2] \la \pa E_{n,\si}/\pa \phi_1
\ra$; however, $\phi_1$ continues to be given by Eq.~\eqref{ii7} and $\la V_1
\ra$ therefore remains equal to $\hbar \al/(2e)$. Since Eq.~\eqref{eav} can
have a range of values depending on $\phi_0$, we see that $I + (e/\hbar)
\sum_{n,\si} [f (E_{n,\si})-1/2] \la \pa E_{n,\si}/\pa \phi_1 \ra$ can have a
range of values while $\la V_1 \ra$ has a fixed value. This corresponds to a
Shapiro plateau in a plot of $\la V_1 \ra$ versus the dc part of $I_1$.

The discussion presented above and the expression in Eq.~\eqref{eav}
imply that the width of the plateau at $(2e/\hbar) \la V_1 \ra = -(m/l) \om$
will be proportional to the Fourier coefficient $c_l$ multiplied by $J_m (la)$.
We thus expect that $c_l$ can give an idea of the plateau widths at integer
multiples of $\om/|l|$. The numerical results presented in
Sec.~\ref{sec:numerics} confirm this expectation.

The above arguments imply that in contrast to the standard
Josephson junctions made out of two wires, junctions of multiple
wires exhibit Shapiro plateaus for drive frequencies for which $2 e
\la V_1 \ra/(\hbar \om)$ is a rational number $m/l$ ( where $m$ and
$l$ are integers). Since there is a rational number lying between
any two rational numbers, there should be a plateau lying between any
two plateaus leading to a a devil's staircase structure~\cite{maiti,shuk}.
However the plateau width quickly goes to zero
and therefore become difficult to see as either $m$ or $l$ becomes
large; this is because $J_m (la) \to 0$ very rapidly as $|m| \to \infty$ for
a fixed value of $la$, and the Fourier coefficient $c_l \to 0$ as $|l| \to
\infty$ for any smooth periodic function $\pa E_{n,\si}/\pa \phi_1$.

\section{Junction of three superconducting wires}
\label{sec:three}

\subsection{Three $s$-wave superconducting wires}

We first consider a system of three $s$-wave SC wires. (We will only consider
spin-up quasiparticles here; the ABS energies, wave functions and Berry
curvature are identical
for spin-down quasiparticles in this case). Equation~\eqref{abs2} implies that
$1/a_\si^2(E_\si)$ is an eigenvalue of $Se^{i\phi}S^*e^{-i\phi}$, and
there must be three such eigenvalues. The $E \to - E$ symmetry implies
that two of the ABS must have energies $\pm E_\si$ (which are generally not
equal to either zero or $\pm \De$), while the third ABS must have $E_\si$ equal
to either zero or $\pm \De$. To see which of these two possibilities occur
for the third ABS, we first consider the trivial case $S=I$. It is then
clear that all the ABS energies lie at $\pm \De$. If we now smoothly move $S$
away from $I$, the $E \to - E$ symmetry implies that two of the ABS
can move away from $\pm \De$ as a pair, but the third state must remain fixed
at $E_\si^2/\De^2 = 1$ (which is not a bound state). We thus conclude in
general that one of the eigenvalues of $Se^{i\phi}S^*e^{-i\phi}$ must be equal
to 1, while the other two eigenvalues form a complex conjugate pair
$a_\si^2(E_\si)$ and $a_\si^{*2} (E_\si)$. Using the fact that
the sum of the eigenvalues of a matrix is equal to its trace, we find that
the energies of two of the ABS are given by~\cite{meyer}
\beq \frac{E_\si}{\De} ~=~ \pm \frac{1}{2} ~\sqrt{ 1 ~+~ {\rm Tr} (S e^{i\phi}
S^* e^{-i\phi})}~. \label{abs5} \eeq
If the matrix $S$ is symmetric, Eq.~\eqref{abs5} implies that two of the
ABS have energies
\bea \frac{E_\si}{\De} &=& \pm [ 1 - |S_{12}|^2 \sin^2 (\frac{\phi_1 - \phi_2}{
2}) - |S_{13}|^2 \sin^2 (\frac{\phi_1 - \phi_3}{2}) \non \\
&& ~~~~~- |S_{23}|^2 \sin^2 (\frac{\phi_2 - \phi_3}{2}) ]^{1/2}.
\label{abs6} \eea

A general parametrization of $3 \times 3$ unitary matrices $S$ has been given
in Ref.~\onlinecite{xie1}. Instead of looking at the most general case,
however, we will first consider a special family of matrices which is
completely symmetric
under all possible permutations of the three wires. Apart from an overall
phase (which is unimportant since it cancels out between $S$ and $S^*$ in
Eq.~\eqref{abs2}), it turns out that such completely symmetric matrices are
labeled by a single real parameter $\lam$ and have the form~\cite{lal}
\bea S &=& \left( \begin{array}{ccc}
r & t & t \\
t & r & t \\
t & t & r \end{array} \right), \non \\
{\rm where} ~~~r &=& -~ \frac{1 ~+~ i \lam}{3 ~+~ i \lam}, \non \\
t &=& \frac{2}{3 ~+~ i \lam}. \label{smat3} \eea
The physical significance of $\lam$ is that it is the strength of a barrier
between the three wires~\cite{lal}. For $\lam=0$, there is no
barrier and the transmission probability $|t|^2 =4/9$ has the
maximum possible value allowed by unitarity for three wires which
are completely symmetric with respect to each other. For
$\lam=\infty$, the barrier is infinitely large and $|t|^2 = 0$.
With $S_{12}=S_{13}=S_{23}=t$, Eq.~\eqref{abs6} gives
\beq \frac{E_\si}{\De} ~=~ \pm ~\left[ \frac{|\sum_{i=1}^3
~e^{i\phi_i}|^2 ~+~ \lam^2}{9 ~+~ \lam^2} \right]^{1/2}. \label{abs7} \eeq
For $\lam =0$, we get $E_\si/\De = \pm
|\sum_{i=1}^3 ~e^{i\phi_i}|/3$, while $\lam \to \pm \infty$ gives
$E_\si/\De \to \pm 1$. To understand the form of $E_\si/\De$ better,
it is useful to study the function \beq F ~=~ \sum_{i=1}^3
~e^{i\phi_i} ~=~ e^{i\phi_1} ~+~ e^{i\phi_2} ~+~ e^{i\phi_3}. \eeq
We can show that $F=0$ if the three phases $\phi_1, ~\phi_2,
~\phi_3$ are $2\pi/3$ apart. For instance, if we set $\phi_3 = 0$,
we get $F=0$ if $(\phi_1,\phi_2)$ is equal to either
$(2\pi/3,4\pi/3)$ or $(4\pi/3,2\pi/3)$. If we expand around one of
these points, say, $\phi_1 = 2\pi/3 +\de \phi_1$ and $\phi_2 =
4\pi/3 +\de \phi_2$, we obtain \beq F ~=~ - ~\frac{\sqrt{3}}{2}
~(\de \phi_1 ~-~ \de \phi_2) ~+~ \frac{i}{2} ~(\de \phi_1 ~+~ \de
\phi_2) \eeq to first order in $\de \phi_1, ~\de \phi_2$.
Equation~\eqref{abs7} then takes the form \beq \frac{E_\si}{\De} = \pm
~\sqrt{\frac{(3/4) (\de \phi_1 - \de \phi_2)^2 + (1/4) (\de \phi_1 +
\de \phi_2)^2 + \lam^2}{9 ~+~ \lam^2}}. \label{abs8} \eeq We can
think of Eq.~\eqref{abs8} as the energy-momentum dispersion of a
particle moving in two dimensions with momentum $k_x = \sqrt{3} (\de
\phi_1 - \de \phi_2)$ and $k_y = \de \phi_1 + \de \phi_2$; the
dispersion has the relativistic form
\bea E_\si^2 &=& v^2 ~(k_x^2 ~+~ k_y)^2 ~+~ m^2, \non \\
{\rm where} ~~~v &=& \frac{\De}{2 ~\sqrt{9 + \lam^2}} \non \\
{\rm and} ~~~m &=& \frac{\De \lam}{\sqrt{9 + \lam^2}} \eea
are the `velocity' and `mass' respectively. The situation described above is
similar to what happens in graphene close to the two Dirac points~\cite{neto};
a mass term can be induced there by applying a sublattice potential or a
spin-orbit interaction.

Turning to the Berry curvature matrix $B_{n,\si}$, we find that
Eqs.~\eqref{bij} and the antisymmetry imply that the matrix is described by
a single real parameter $b_{n,\si}$ as
\beq B_{n,\si} ~=~ \left( \begin{array}{ccc}
0 & b_{n,\si} & -b_{n,\si} \\
-b_{n,\si} & 0 & b_{n,\si} \\
b_{n,\si} & -b_{n,\si} & 0 \end{array} \right). \label{berry3} \eeq
The value of $b_{n,\si}$ depends on the $\phi_i$'s, $S$ and the particular ABS
band $n$ and spin $\si$ which is being considered. Typically, we find that the
Berry curvature is large near those values of the $\phi_i$'s where two of the
ABS are almost degenerate in energy.

\begin{figure}[h]
\subfigure[]{\includegraphics[width=8.6cm]{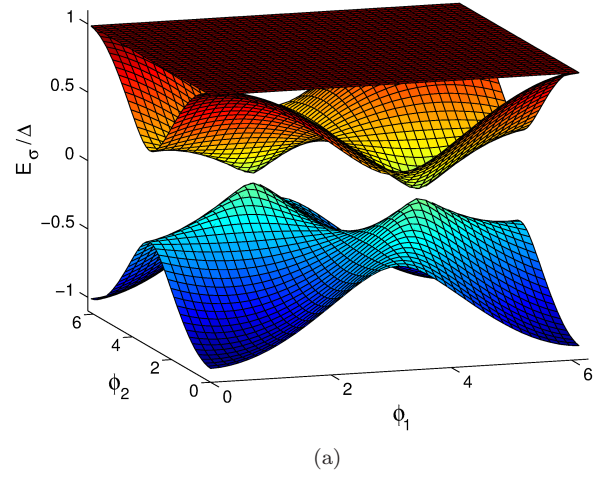}} \\
\subfigure[]{\includegraphics[width=8.6cm]{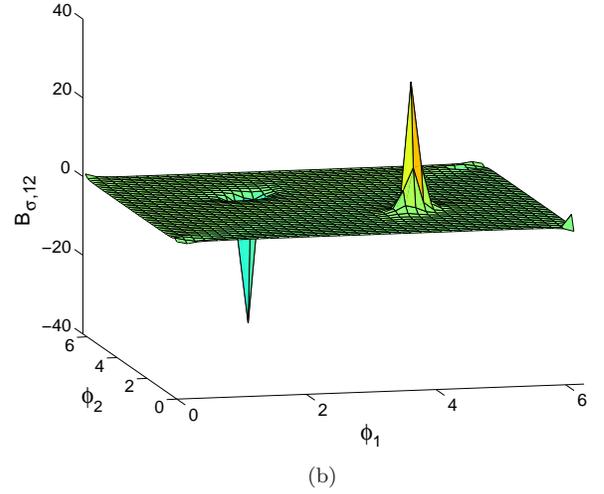}}
\caption{(a) Surface plot of ABS energies (in units of $\De$) vs $(\phi_1,
\phi_2)$ for the $S$-matrix given in Eq.~\eqref{smat3} with $\lam=0.1$. For all
values of $(\phi_1,\phi_2)$, one of the energies lies at $E_\si/\De =1$ and the
other two ABS energies appear as a $\pm E_\si$ pair; the gap between these two
bands is minimum at $(2\pi/3,4\pi/3)$ and $(4\pi/3,2\pi/3)$. (b) Surface plot
of Berry curvature $B_{\si,12}$ vs $(\phi_1,\phi_2)$ for the ABS band with
$E_\si/ \De < 0$. There are peaks at $(2\pi/3, 4\pi/3)$ and $(4\pi/3,2\pi/3)$
with negative and positive signs respectively, and $Ch_{\si,12}=0$.}
\label{fig03} \end{figure}

In the rest of this section, we will set $\phi_3 = 0$ for
convenience. For the $S$-matrix given in Eq.~\eqref{smat3} and
$|\lam| \ll 1$, we numerically find the following results in the
$(\phi_1,\phi_2)$ plane. The parameter $B_{n,\si,12} = b_{n,\si}$ is
peaked near the two points $(\phi_1,\phi_2) = (2\pi/3,4\pi/3)$ and
$(4\pi/3,2\pi/3)$ and is close to zero everywhere else; the peaks
occur where two of the ABS have energies close to zero and are
therefore almost degenerate. Furthermore, for the ABS with one of
the energies given in Eq.~\eqref{abs8}, the sign of the peak is
given by $sgn (\lam E_\si)$ at the first point and $-sgn (\lam
E_{n,\si})$ at the second point. In all cases, we find that the
Chern number $Ch_{n,\si,12}$ defined in Eq.~\eqref{c12} is equal to
zero as the contributions from the two peaks cancel each other.
Figure~\ref{fig03} shows surface plots versus $(\phi_1,\phi_2)$ of
(a) the three bands of ABS energies and (b) the Berry curvature
$B_{n,\si,12}$ for the ABS band with $E_{n,\si}/\De < 0$, for the
$S$-matrix given in Eq.~\eqref{smat3} with $\lam=0.1$. (We have set
$\phi_3 = 0$). We see that $B_{n,\si,12}$ has peaks at
$(2\pi/3,4\pi/3)$ and $(4\pi/3,2\pi/3)$ with negative and positive
signs respectively, and the Chern number $Ch_{n,\si,12}$ is equal to
zero. The fact that $Ch_{n,\si,12}=0$ is generally true if $S$ is
symmetric, i.e., time-reversal symmetric~\cite{xie1}.

\begin{figure}[h]
\subfigure[]{\includegraphics[width=8.6cm]{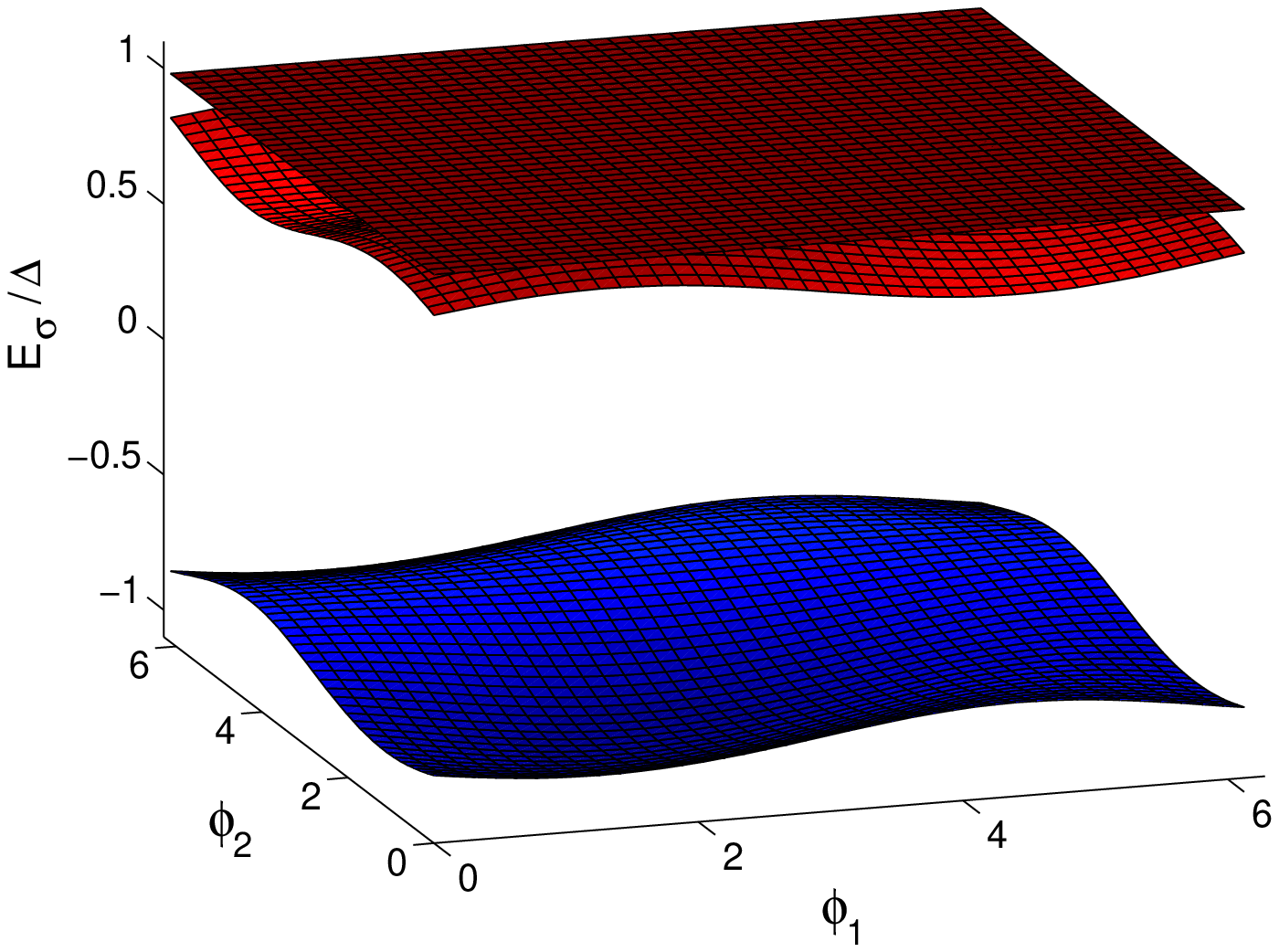}} \\
\subfigure[]{\includegraphics[width=8.6cm]{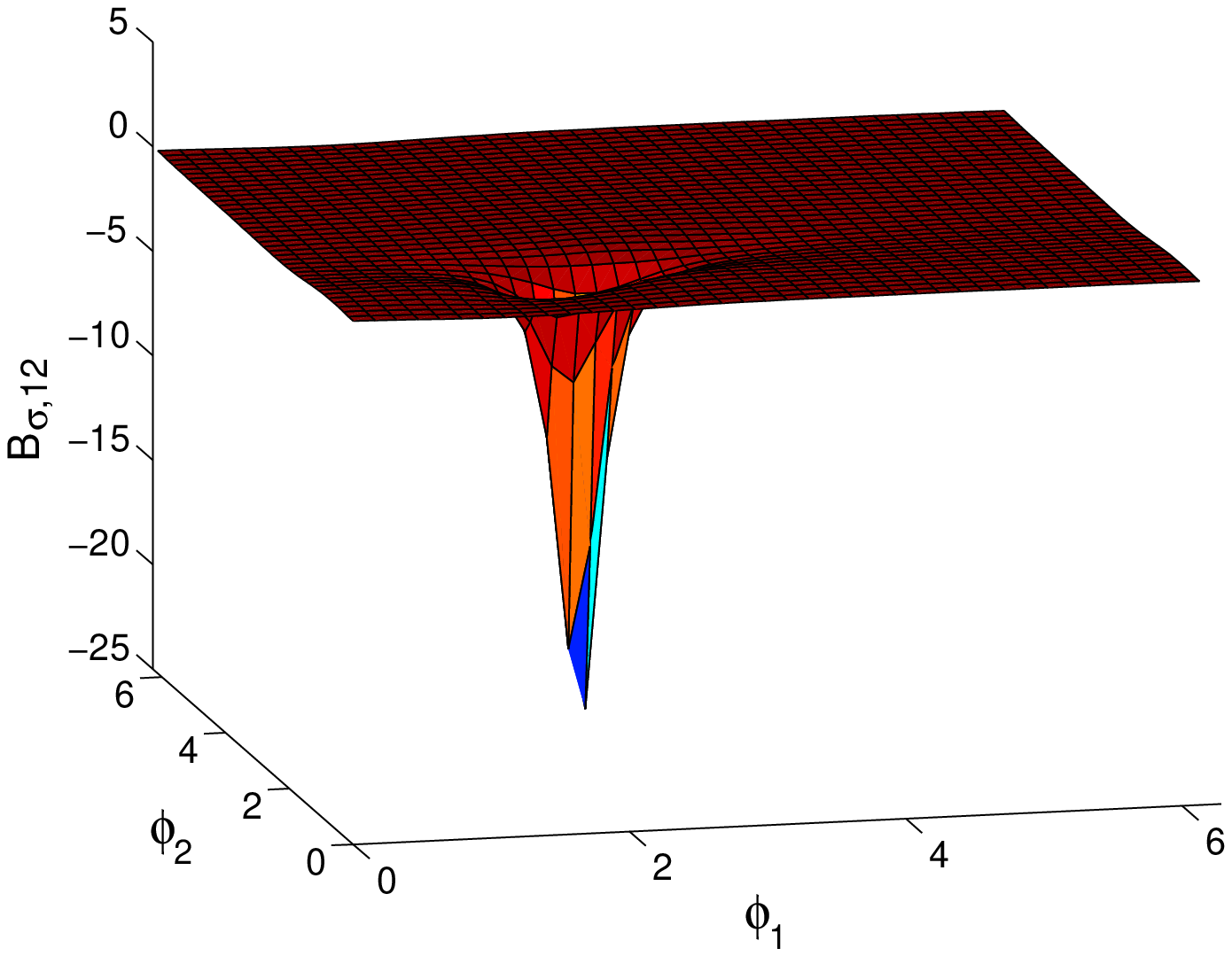}}
\caption{(a) Surface plot of ABS energies (in units of $\De$) vs $(\phi_1,
\phi_2)$ for the $S$-matrix given in Eq.~\eqref{srand}. For all values of
$(\phi_1,\phi_2)$, one of the energies lies at $E_\si/\De = 1$ and the other
two ABS energies appear as a $\pm E_\si$ pair. The gap between the band
with $E_\si/\De = 1$ and the band with $0 < E_\si/\De < 1$ is minimum at
$(2.04,1.57)$. (b) Surface plot of Berry curvature $B_{\si,12}$ vs
$(\phi_1,\phi_2)$ for the $S$-matrix given in Eq.~\eqref{srand} and the ABS
with $E_\si/\De < 0$. There is a single peak with a negative sign at
$(2.04,1.57)$, and $Ch_{\si,12}=-1$.} \label{fig04} \end{figure}

If $S$ is not a symmetric matrix the Chern number $Ch_{n,\si,12}$ can be
non-zero~\cite{xie1}. In that case we typically find that there is a single
peak in $B_{n,\si,12}$ and $Ch_{n,\si,12} =\pm 1$. Furthermore, the peak in
$B_{n,\si,12}$
usually occurs at a point in the $(\phi_1,\phi_2)$ plane where one of the ABS
is almost degenerate with the ABS which always lies at $E_\si^2/\De^2 = 1$.
Figure~\ref{fig04} shows surface plots versus $(\phi_1,\phi_2)$ of (a) the
three bands of ABS energies and (b) the Berry curvature versus $(\phi_1,
\phi_2)$ for the ABS band with $E_\si/\De < 0$, for a randomly generated
$S$-matrix given by
\begin{widetext}
\beq S ~=~ \left( \begin{array}{ccc}
0.8389 - 0.0346i & -0.2399 + 0.3146i & -0.3542 + 0.1138i \\
-0.0163 + 0.3446i & 0.7254 - 0.0341i & -0.4820 + 0.3483i \\
0.2591 + 0.3299i & 0.3589 + 0.4328i & 0.7119 - 0.0341i \end{array} \right).
\label{srand} \eeq
\end{widetext}
(We again set $\phi_3 = 0$). We see that the Berry curvature has a peak with a
negative sign at $(\phi_1,\phi_2) =(2.04,1.57)$, and $Ch_{\si,12}$ is equal to
$-1$.

{\bf Large $\lam$ limit of the symmetric $S$-matrix:} It is
interesting to consider what happens if we take the parameter $\lam$
to be large in Eq.~\eqref{smat3}; as mentioned there, this
corresponds to having a large barrier at the junction between the
three wires. Keeping terms only up to order $1/\lam$, we find that
$r \simeq - 1 - 2i/\lam$ and $t \simeq - 2i /\lam$. The operator in
Eq.~\eqref{abs2} then takes the form
\bea && S e^{i \phi} S^* e^{-i\phi} ~\simeq~ I_3 ~+~ i M, \non \\
&& M = \frac{2}{\lam} \left( \begin{array}{ccc}
0 & 1 - e^{i(\phi_1 - \phi_2)} & 1 - e^{i(\phi_1 - \phi_3)} \\
1 - e^{i(\phi_2 - \phi_1)} & 0 & 1 - e^{i(\phi_2 - \phi_3)} \\
1 - e^{i(\phi_3 - \phi_1)} & 1- e^{i(\phi_3 - \phi_2)} & 0 \end{array} \right),
\non \\
&& \label{matm} \eea
where $I_3$ is the $3 \times 3$ identity matrix and $M$ is a Hermitian
matrix. The eigenvalues of $M$ turn out to be zero and
\beq \pm 2 ~\sqrt{\sin^2 (\frac{\phi_1 - \phi_2}{2}) + \sin^2 (\frac{\phi_1
- \phi_3}{2}) + \sin^2 (\frac{\phi_2 - \phi_3}{2})}. \label{eigm} \eeq
It follows from Eqs.~\eqref{abs2}, \eqref{matm} and \eqref{eigm} that one of
the ABS energies lies at $E_\si^2/\De^2 = 1$, while the other two are given by
\bea \frac{E_\si}{\De} &\simeq& \pm [1 ~-~ \frac{2}{\lam^2} ~\{\sin^2
(\frac{\phi_1 - \phi_2}{2}) ~+~ \sin^2 (\frac{\phi_1 - \phi_3}{2}) \non \\
&& ~~~~~~~~~~~~~~~~+~ \sin^2 (\frac{\phi_2 - \phi_3}{2})\}]
\label{largelam} \eea up to terms of order $1/\lam^2$. For the ABS
with negative energy, Eq.~\eqref{largelam} implies that \beq
\frac{\pa E_\si}{\pa \phi_1} ~\simeq~ \frac{\De}{\lam^2} ~[\sin
(\phi_1 - \phi_2) ~+~ \sin (\phi_1 - \phi_3)]. \label{dedphi} \eeq
We thus see that the contribution of this term to the current (in
Eqs.~\eqref{curr2} and \eqref{ii6} for instance) scales as
$\De/\lam^2$ for large $\lam$. This observation will be useful
later.

It is clear from Eqs.~\eqref{abs2} and \eqref{matm} that in the
large $\lam$ limit, the ABS wave functions and hence the Berry
curvature depend only on the phases $\phi_i$ and not on $\lam$. The
Berry curvature is large near the point $\phi_1 = \phi_2 = \phi_3$
since the three eigenvalues of $M$ are degenerate there. However we
find that the peak value of the Berry curvature is much smaller here
compared to its value for small $\lam$ as shown in Fig.~\ref{fig03}.

\subsection{Systems with some $p$-wave superconducting wires}

We first consider a system in which all the three SC wires have $p$-wave
pairing. The diagonal matrix $\eta$ defined after Eq.~\eqref{pwave} is then
equal to $-I$, and the ABS energies will be given by the eigenvalue equation
\beq S e^{i \phi} S^* e^{-i\phi} \psi_\si (E_\si) ~=~ -~ \frac{1}{a_\si^2
(E_\si)} \psi_\si (E_\si) \label{abs13} \eeq
for both spin-up and -down quasiparticles; hence each of the energies will
have a two-fold degeneracy. Then an argument similar to the one presented at
the beginning of Sec.~\ref{sec:three} A will show that one of the ABS energies
will always lie at $E_\si=0$ (corresponding to $a_\si^2 (E_\si) = -1$), while
the other two energies must be of the form $\pm E_\si$. Then the fact that
the sum of the eigenvalues of a matrix is equal to its trace implies that
the energies of two of the ABS are given by
\beq \frac{E_\si}{\De} ~=~ \pm \frac{1}{2} ~\sqrt{ 3 ~-~ {\rm Tr} (S e^{i\phi}
S^* e^{-i\phi})}~. \label{abs14} \eeq

We next consider a system in which two of the SC wires, say 1 and 2, are
$p$-wave while wire 3 is $s$-wave. Then the diagonal matrix $\eta$ will have
entries given by $(-1,-1,1)$, and the ABS energies will be given by
Eq.~\eqref{abs4u} for spin-up quasiparticles. (As discussed after
Eq.~\eqref{absx2}, the ABS energies for spin-down quasiparticles will be given
by $-1$ times the spin-up energies). We now find that none of the eigenvalues
$1/a_\si^2 (E_\si)$ are equal to $\pm 1$ in general, and we no longer have
simple expressions like Eq.~\eqref{abs5} or \eqref{abs14} for the energies.
However, we discover an interesting fact if $S$ is a symmetric matrix. If
$\phi_1 = \phi_2$ and $\phi_3$ takes any value, one of the ABS energies lies
at $E_\si=0$ corresponding to $a_\si^2 (E_\si) = -1$.
Thus, if $\phi_3$ is held fixed, there is a line in the $(\phi_1,\phi_2)$
plane on which one of the ABS energies is equal to zero.

\section{Numerical results for three superconducting wires}
\label{sec:numerics}

We will now present our numerical results. When calculating the
currents $I_i$, we have to sum over spin-up and -down
quasiparticles and also over the three bands of ABS energies given
by functions of $\phi_1$ and $\phi_2$ (we will generally set $\phi_3
= 0$). As usual we will denote the ABS energies by $E_{n,\si}$,
where $n=1,2,3$ labels the bands and $\si = \pm 1$ labels spin-up
and spin-down quasiparticles. If $I_{i,n,\si}$ is the contribution
to the current in wire $i$ from band $n$ and spin $\si$, the total
current in wire $i$ is given by~\cite{riwar}
\bea I_i &=& \frac{1}{2}~ \sum_{n,\si} ~I_{i,n,\si} ~[f (E_{n,\si}) ~-~
\frac{1}{2}]. \label{iins} \eea
In general, the SC phases $\phi_i$ and the ABS
energies $E_{n,\si}$ can vary with time. We will assume that Eq.~\eqref{iins}
is valid at all times with $E_{n,\si}$ being the instantaneous energy.

In the numerical calculations presented below, we have chosen the SC
gap to be $\De = 10^{-6}$ eV in order to obtain experimentally
reasonable values of the currents $I_i$ (of the order of nA) and the
frequency $\om$ (of the order of GHz) used to study Shapiro
plateaus. Another reason for choosing $\De = 10^{-6}$ eV, along with
appropriate values of the elements of the $S$ matrix, is to show
large variations and striking peaks in the ABS energies and Berry
curvature in Fig.~\ref{fig03}. We note, however,
that the value of $\De = 10^{-6}$ eV is much smaller than typical
experimental values of the order of $10^{-3}$ eV. As stated after
Eq.~\eqref{dedphi}, if we scale $\De$ up from $10^{-6}$ to $10^{-3}$
eV and $\lam^2$ up by the same factor of $10^3$, i.e., scale $\lam$
up by a factor of about 30, the values of the currents will not
change drastically (the currents will change to some extent because
the Berry curvature changes as we vary $\lam$ from small to large
values). To conclude, we will present numerical results for $\De =
10^{-6}$ eV so as to illustrate various ideas (such as
transconductance, Shapiro plateaus and junction magnetic moment)
most clearly, with the understanding that the results will not
change qualitatively if we use more realistic values of $\De$ along
with suitable values of $S$. We also note that in what follows, we will 
only consider the contribution to the currents from the Andreev bound states. 

\subsection{Three $s$-wave superconducting wires}

To begin, we will examine the case of three $s$-wave SC wires. In
this case, the energies, wave functions and Berry curvature are
identical for spin-up and -down quasiparticles. We will
therefore consider only spin-up quasiparticles in the following.
Next, we know that one of the ABS energy bands always lies at the
edge between the gap and the bulk ($E_\si^2/\De^2 = 1$); hence it is
not really a bound state. We will therefore not consider this band.
The other two bands have energies $\pm E_\si$. We will only consider
cases where these energies are gapped away from zero. At
temperatures much lower than the gap, the band with positive
(negative) energy will be unoccupied (occupied) and will therefore
have $f (E_{n,\si}) - 1/2$ equal to $-1/2 ~(1/2)$ respectively.
Since the energies come in $\pm E$ pairs, we have $\sum_{n=\pm}
E_{n,\si} [f (E_{n,\si}) - 1/2] = E_{-,\si}$, where $E_{\pm,\si}$
denotes the positive (negative) energy values for spin $\si$. For the
Berry curvature, we have $\sum_{n=\pm} B_{n,\si,12} [f (E_{n,\si}) -
1/2] \simeq (B_{-,\si,12} - B_{+,\si,12})/2$, where $B_{\pm,\si,12}$
denotes the Berry curvatures in the positive (negative) energy bands.
(We find numerically that for each value of the $\phi_i$'s,
$B_{+,\si,12}$ and $B_{-,\si,12}$ have almost the same magnitude but
opposite signs, as mentioned after Eq.~\eqref{bsym}). We also note
that $E_{\pm,\si}$ and $B_{\pm,\si,12}$ have the same values for
$\si = \pm 1$. Hence, in many of the equations below, we will write
only $\si = +1$ and include a factor of $2$ for spin.

\begin{figure}[h]
\epsfig{figure=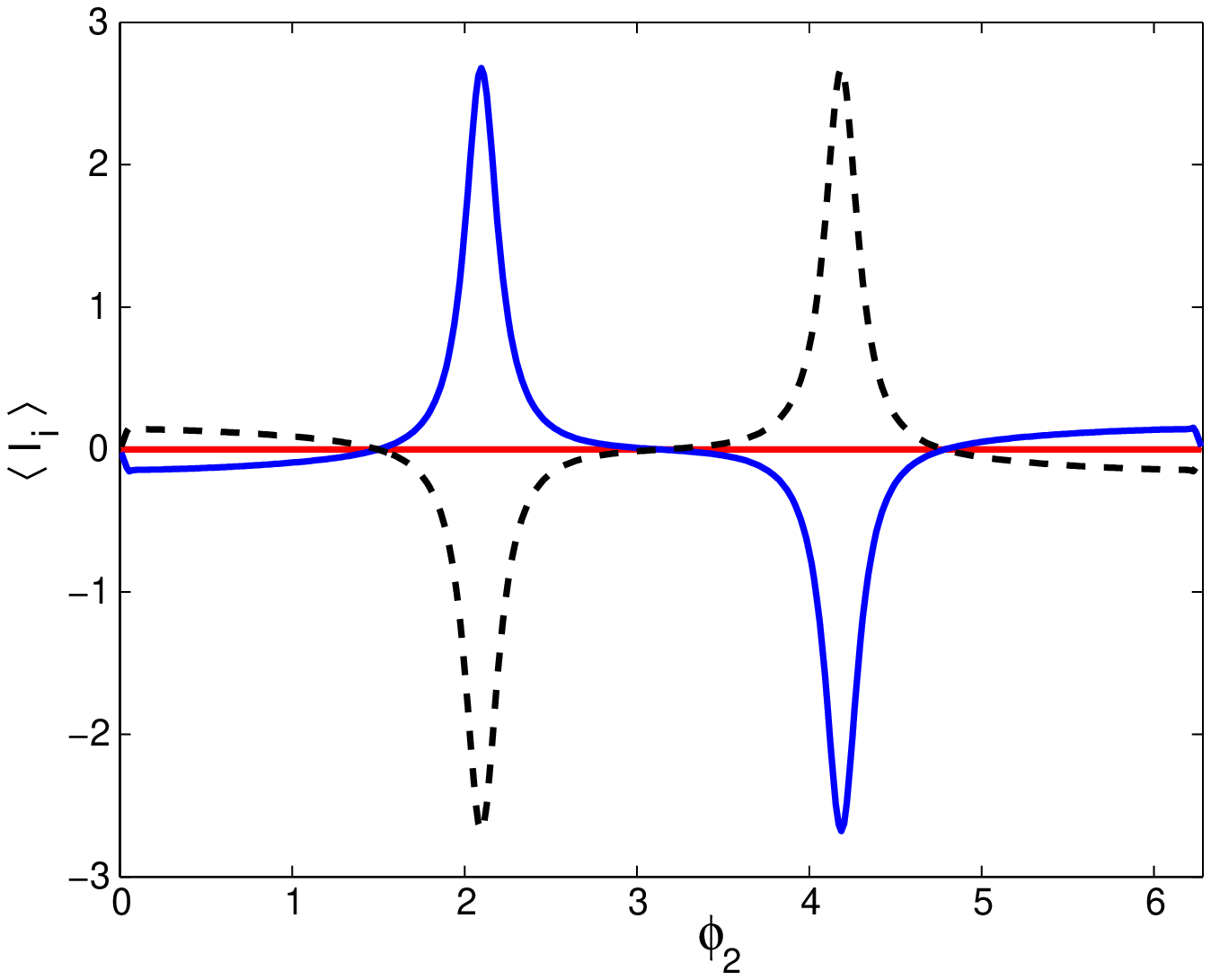,width=8.6cm} \caption{Plot of average currents
$\la I_1 \ra$ (red solid line which lies at zero), $\la I_2 \ra$ (blue solid
line), and $\la I_3 \ra$ (black dashed line) vs $\phi_2$ lying in the range
$[0,2\pi]$. We have chosen $V_1 = 5 \times 10^{-5}$ V, $V_2 = V_3 = 0$,
$\phi_3 = 0$, $\De = 10^{-6}$ eV, and the $S$-matrix has the form given in
Eq.~\eqref{smat3} with $\lam=0.1$. The currents are in units of $10^{-4}
\times (e/\hbar) ({\rm eV}) \simeq 24$ nA.} \label{fig05} \end{figure}

We first look at the ac Josephson effect, namely, the case of constant
voltages $V_i$; this is discussed in Eq.~\eqref{curr2}, but we have to
multiply the expression in that equation by 2 to account for spin.
In Fig.~\ref{fig05}, we show the average currents $\la I_1 \ra$, $\la I_2 \ra$
and $\la I_3 \ra$ as functions of $\phi_2$ for a system with $V_1 = 5 \times
10^{-5}$ V (so that $2eV_1 = 10^{-4}$ eV), $V_2 = V_3 = 0$, $\phi_3 = 0$,
$\De = 10^{-6}$ eV, and an $S$-matrix of the form
given in Eq.~\eqref{smat3} with $\lam=0.1$. (As described in Eq.~\eqref{curr3},
we will take the units of current to be $e/\hbar$ times eV which is
equal to $(2\pi)V /(h/e^2) ~\simeq 2.43 \times 10^{-4}$ A). We can understand
the main features of Fig.~\ref{fig05} as follows. First, as noted after
Eq.~\eqref{int1}, $\la I_1 \ra = 0$; hence current conservation implies that
$\la I_2 \ra = - \la I_3 \ra$. Second, we find numerically that the entire
contribution to $\la I_2 \ra$ and $\la I_3 \ra$ comes from the Berry curvature
term, i.e., the second term on the right hand side of Eq.~\eqref{qt}. Since
$V_1$ is constant, $\phi_1 = (2e/\hbar) V_1 t$ varies linearly with time and
covers the full range of $2\pi$ in a time $T= 2\pi \hbar/(2eV_1)$.
Equations~(\ref{qt}-\ref{qit1}) then imply that $\la I_1 \ra = 0$, and
\bea \la I_2 \ra &=& - \la I_3 \ra ~=~ \frac{2e^2 V_1}{\hbar} ~\int_0^{2\pi} ~
\frac{d\phi_1}{2\pi} \non \\
&& \times ~[B_{-,+1,12} (\phi_1,\phi_2) ~-~ B_{+,+1,12} (\phi_1,\phi_2)].
\label{i123} \eea
Since the Berry curvature
$B_{-,+1,12}$ has large values around $(\phi_1,\phi_2)=(2\pi/3,4\pi/3)$ with
a negative sign and around $(4\pi/3,2\pi/3)$ with a positive sign (as shown
in Fig.~\ref{fig03}), we see from Eq.~\eqref{i123} that, for
$2eV_1 =10^{-4}$ eV, $\la I_2 \ra$ will be large
and positive around $\phi_2 = 2\pi/3$ (with $I_2 (t)$ getting its maximum
contribution at the time when $\phi_1$ passes through $4\pi/3$) and negative
around $\phi_2 = 4\pi/3$ (with the maximum contribution to $I_2 (t)$ coming
from the time when $\phi_1$ passes through $2\pi/3$). This explains the
locations and signs of the peaks in $\la I_2 \ra$ in Fig.~\ref{fig05}.
In addition, we have verified numerically that Eq.~\eqref{i2c12} is satisfied
with $Ch_{-,+1,12} = Ch_{+,+1,12} = 0$.

For $2eV_1 = 10^{-4}$ eV, Eq.~\eqref{i123} and the fact that $B_{+,+1,12}
(\phi_1, \phi_2) \simeq - B_{-,+1,12} (\phi_1,\phi_2)$ imply that as a
function of $\phi_2$, $\la I_2 \ra = - \la I_3 \ra$ is equal, in units of
$10^{-4} \times (e/\hbar) ({\rm eV}) \simeq 24$ nA,
to twice the average Berry curvature which is defined as
\beq {\bar B}_{12} (\phi_2) ~=~ \int_0^{2\pi} ~\frac{d\phi_1}{2\pi} ~
B_{-,+1,12} (\phi_1,\phi_2), \label{barb} \eeq
For instance, we see in Fig.~\ref{fig05} that $2 {\bar B}_{12} \simeq 2.68$ for
$\phi_2 = 2\pi/3$ and $\phi_3 = 0$; this value will be used later.

\begin{figure}[h]
\epsfig{figure=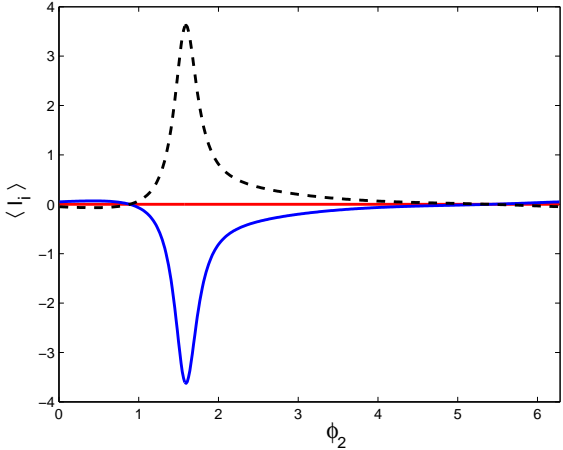,width=8.6cm}
\caption{Plot of average currents $\la I_1 \ra$ (red solid line which lies at
zero), $\la I_2 \ra$ (blue solid line), and $\la I_3 \ra$ (black dashed line)
vs $\phi_2$ lying in the range $[0,2\pi]$. We have chosen $V_1 = 5 \times
10^{-5}$ V, $V_2 = V_3 = 0$, $\phi_3 = 0$, $\De = 10^{-6}$ eV, and the
$S$-matrix has the form given in Eq.~\eqref{srand}. The currents are in units
of $10^{-4} \times (e/\hbar) ({\rm eV}) \simeq 24$ nA.}
\label{fig06} \end{figure}

Similarly, in Fig.~\ref{fig06}, we show $\la I_i \ra$ as functions of $\phi_2$
for a system with $V_1 = 5 \times 10^{-5}$ V (so that $2eV_1 = 10^{-4}$ eV),
$V_2 = V_3 = 0$, $\phi_3 = 0$, $\De = 10^{-6}$ eV, and an $S$-matrix of the
form given in Eq.~\eqref{srand}. Once again we find that $\la I_1 \ra = 0$,
$\la I_2 \ra = - \la I_3 \ra$, and the entire contribution to $\la I_2
\ra$ and $\la I_3 \ra$ comes from the Berry curvature term. Furthermore, we see
only a single peak in $\la I_2 \ra$ with a negative sign; the peak is located
at $\phi_2 = 1.57$ which is consistent with the sign and location of the peak
in the Berry curvature shown in Fig.~\ref{fig04}. Once again, we have
confirmed numerically that Eq.~\eqref{i2c12} holds with $Ch_{-,+1,12} =
- Ch_{+,+1,12} = -1$.

We next look for Shapiro plateaus in an $RC$ circuit involving three wires;
we will assume that the all resistances (capacitances) are equal to $R$ ($C$).
As discussed in Eq.~\eqref{i1}, we consider a case where $I_1 = I + A \sin
(\om t)$, and $V_2 = V_3 = 0$ so that $\phi_2$ and $\phi_3$ are constant
in time. Using Eqs.~\eqref{ii3} and \eqref{ii6} we then obtain the following
equations
\bea {\ddot \phi}_1 + \ga {\dot \phi}_1 &+& \frac{2e^2}{\hbar^2 C}
\frac{\pa E_{-,+1}}{\pa \phi_1} = \frac{e}{\hbar C} ~[I + A \sin (\om t)],
\label{ii81} \\
I_2 ~-~ I_3 &=& \frac{2e}{\hbar} (\frac{\pa E_{-,+1}}{\pa \phi_2} ~-~
\frac{\pa E_{-,+1}}{\pa \phi_3}) \non \\
&& +~ 2e (B_{-,+1,12} ~-~ B_{+,+1,12}) {\dot \phi}_1, \label{ii82} \\
I_2 ~+~ I_3 &=& - ~I_1 ~=~ - I ~-~ A \sin (\om t), \label{ii83} \eea
where we have included a factor of 2 for spin. We can solve these 
equations numerically. Equation~\eqref{ii81} does not involve
the Berry curvature and can be solved without using the next two equations.
Equation~\eqref{ii82} involves the Berry curvature and ${\dot \phi}_1$ which 
can be found using Eq.~\eqref{ii81}. Equation~\eqref{ii83} has a trivial form. 
After solving Eqs.~(\ref{ii81}-\ref{ii82}) over a long time $T$, we can 
calculate the average values
\bea \la V_1 \ra &=& \frac{1}{T} ~\int_0^T ~dt V_1 (t) ~=~ \frac{\hbar}{2e} ~
\frac{\phi_1 (T) ~-~ \phi_1 (0)}{T}, \non \\
\la I_2 - I_3 \ra &=& \frac{1}{T} ~\int_0^T ~dt ~(I_2 (t) ~-~ I_3 (t)).
\label{v1i23} \eea
We can then plot these average values versus $I$ for a particular set of
values of $A$, $\om$, $\phi_2$ and $\phi_3$.

To fix the values of the various parameters, it is convenient to re-define
time in dimensionless units of $\om t$; ${\dot \phi}_1$ and ${\ddot \phi}_1$
are then dimensionless. Next, we define the dimensionless quantities
\beq \al_1 ~=~ \frac{\hbar \om C}{e^2} ~~~~{\rm and}~~~~ \al_2 ~=~
\frac{\hbar}{e^2 R}. \label{al12} \eeq
Equations~(\ref{ii81}-\ref{ii82}) can then be rewritten as
\bea && \frac{\hbar \om}{2} [\al_1 {\ddot \phi}_1 + \al_2 {\dot \phi}_1] +
\frac{\pa E_{-,+1}}{\pa \phi_1} = \frac{\hbar}{2e} ~[I + A \sin t],
\label{ii91} \\
&& \frac{\hbar}{2e} ~(I_2 - I_3) = (\frac{\pa E_{-,+1}}{\pa \phi_2} -
\frac{\pa E_{-,+1}}{ \pa \phi_3}) \non \\
&& ~~~~~~~~~~~~~~~~~~~ + \hbar \om (B_{-,+1,12} - B_{+,+1,12})
{\dot \phi}_1. \label{ii92} \eea
All the terms in Eqs.~(\ref{ii91}-\ref{ii92}) have the dimensions of energy.
In addition, we have $V_1 = (\hbar \om/2e) {\dot \phi}_1$ in units of energy;
hence
\beq \la V_1 \ra ~=~ \frac{\hbar \om}{2e} ~\la {\dot \phi}_1 \ra. \label{v1}
\eeq

\begin{figure}[h]
\epsfig{figure=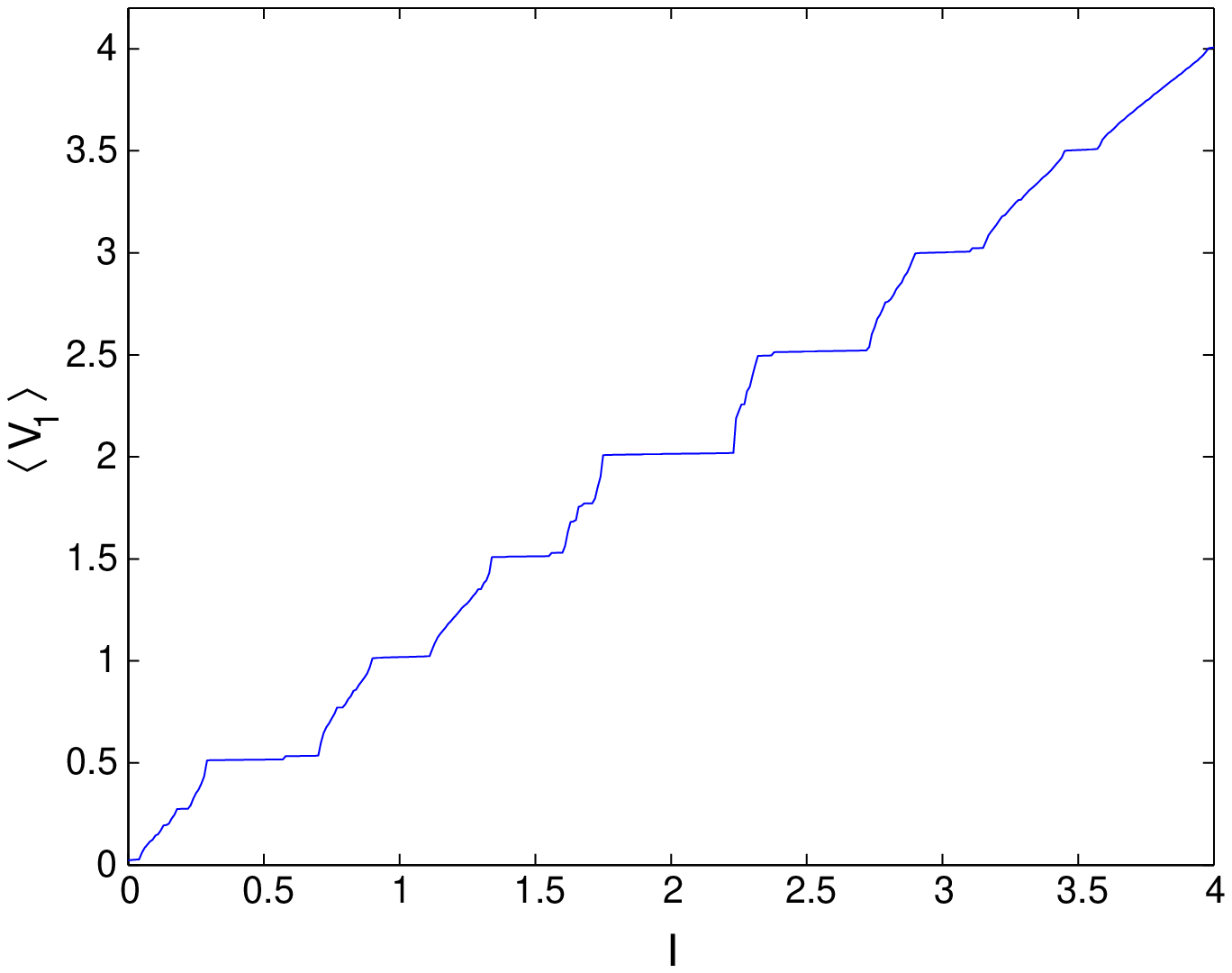,width=8.6cm}
\caption{Plot of $\la V_1 \ra$ vs $I$. We have chosen $\hbar \om = 10^{-6}$ eV,
$A = 4$, $\al_1 = \al_2 = 0.5$, $\phi_2 = 2\pi/3$, $\phi_3 = 0$, and $\De =
10^{-6}$ eV. ($I$ and $A$ are in units of $10^{-6} \times (e/\hbar)
({\rm eV}) \simeq 0.24$ nA, and $V_1$ is in units of $10^{-6}$ V).
The $S$-matrix has the form given in Eq.~\eqref{smat3} with $\lam=0.1$.}
\label{fig07} \end{figure}

\begin{figure}[h]
\epsfig{figure=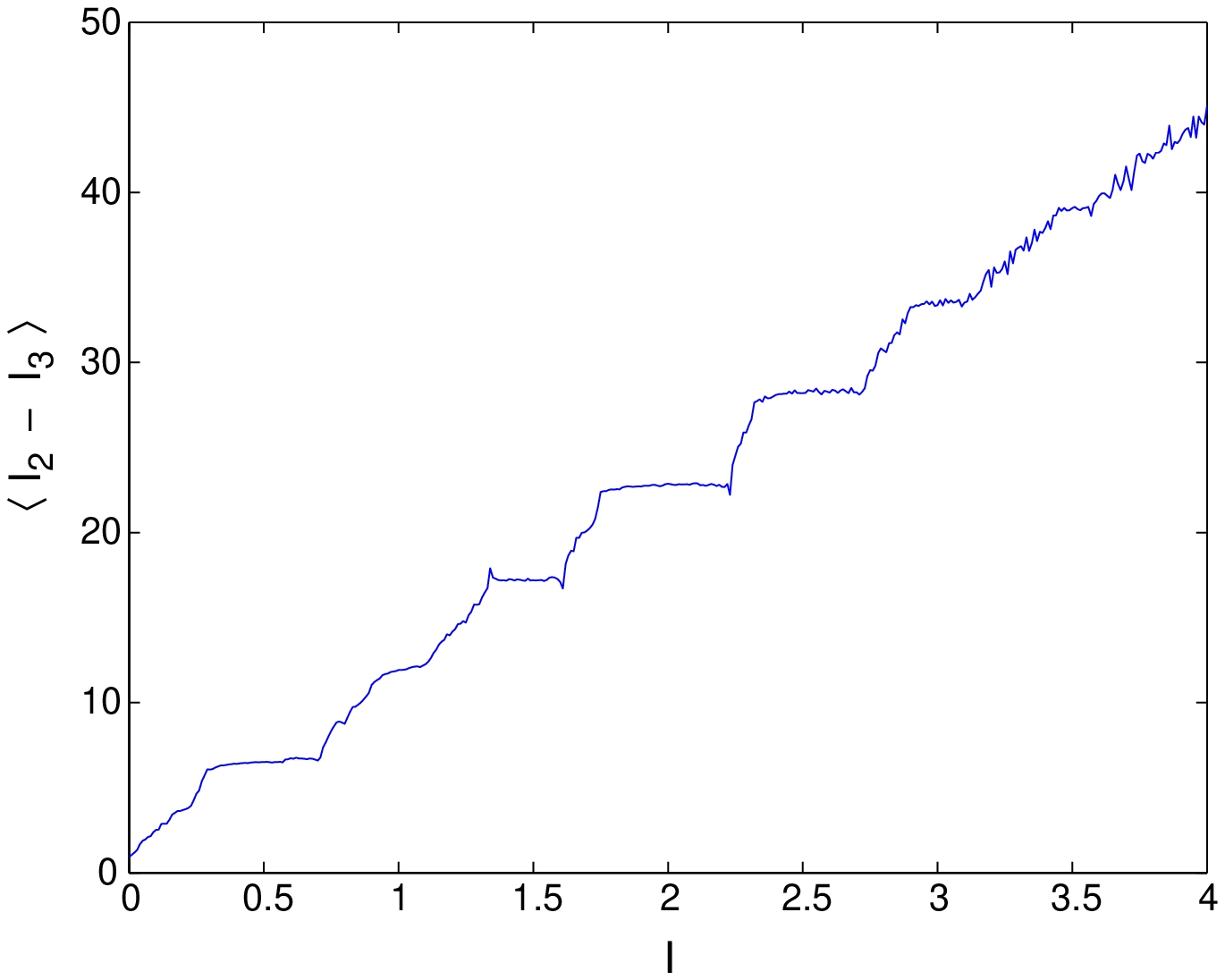,width=8.6cm}
\caption{Plot of $\la I_2 - I_3 \ra$ vs $I$, both in units of $10^{-6} \times
(e/\hbar) ({\rm eV}) \simeq 0.24$ nA. The system parameters are the same as
in Fig.~\ref{fig07}.} \label{fig08} \end{figure}

\begin{figure}[h]
\epsfig{figure=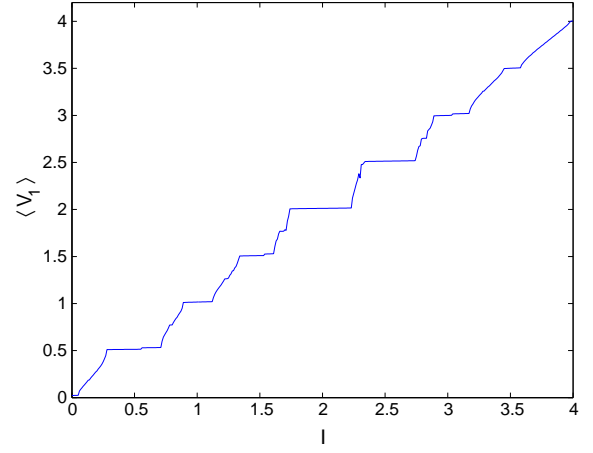,width=8.6cm}
\caption{Plot of $\la V_1 \ra$ vs $I$. We have chosen $\hbar \om = 10^{-6}$ eV,
$A = 4$, $\al_1 = \al_2 = 0.5$, $\phi_2 = \phi_3 = 0$, and $\De = 10^{-6}$ eV.
($I$ and $A$ are in units of $10^{-6} \times (e/\hbar) ({\rm eV}) \simeq 0.24$
nA, and $V_1$ is in units of $10^{-6}$ V). The $S$-matrix has the form given
in Eq.~\eqref{smat3} with $\lam=0.1$.} \label{fig09} \end{figure}

In Fig.~\ref{fig07}, we show a plot of $\la V_1 \ra$ vs $I$ for
$\hbar \om = 10^{-6}$ eV, $A = 4$, $\al_1 = \al_2 = 0.5$, $\phi_2 =
2\pi/3$, $\phi_3 = 0$, and $\De = 10^{-6}$ eV. Note that $\hbar \om
= 10^{-6}$ eV corresponds to $\om \simeq 1.52$ GHz. Furthermore,
Eq.~\eqref{al12} and $\al_1 = \al_2 = 0.5$ imply that $C \simeq 0.080$
pF and $R \simeq 8.22 ~{\rm k}\Omega$. We have taken the $S$-matrix
to be of the form given in Eq.~\eqref{smat3} with $\lam=0.1$. There
are prominent plateaus at $\la V_1 \ra = 0.5, ~1, ~1.5$ 2, $2.5$, 3
and $3.5$ times $10^{-6}$ V (corresponding to integer multiples of
$\hbar \om/ (2e)$) and narrower plateaus at subharmonic values given
by $\la V_1 \ra = 0.25, ~0.75, ~1.75$ and $2.25$ times $10^{-6}$ V
(namely, odd integer multiples of $\hbar \om/(4e) = 0.25$ V).
Figure~\ref{fig08} shows a plot of $\la I_2 - I_3 \ra$ vs $I$ for the
same system parameters; we see plateau-like features around the same
values of $I$ as in Fig.~\ref{fig07}. In Fig.~\ref{fig09}, we show a
plot of $\la V_1 \ra$ vs $I$ for the same system parameters as in
Fig.~\ref{fig07} except that we have chosen $\phi_2 = \phi_3 = 0$.
The plots in Figs.~\ref{fig07} and \ref{fig09} look similar, except
that the subharmonic plateaus at odd-integer multiples of $\hbar
\om/4e$ are somewhat less prominent in Fig.~\ref{fig09}. We note
that the choices of $\phi_2$ and $\phi_3$ are such that the system
passes through a region of large Berry curvature in Fig.~\ref{fig07}
but not in Fig.~\ref{fig09} (see Fig.~\ref{fig03}); however this
makes very little difference to the behavior of $\la V_1 \ra$ since
this quantity is obtained from Eq.~\eqref{ii91} which does not
contain the Berry curvature. On the other hand, we find numerically
that the quantity $\la I_2 - I_3 \ra$ shown in Fig.~\ref{fig08}
mainly gets a contribution from the Berry curvature term in
Eq.~\eqref{ii92}; the first term in that equation (of the form $\pa
E/\pa \phi_i$) makes only a small contribution. For the parameters
chosen in Fig.~\ref{fig09}, we have $\phi_2 = \phi_3$; hence there
is perfect symmetry between wires 2 and 3. In this case, therefore,
we have $\la I_2 - I_3 \ra = 0$ for all values of $I$.

We can relate the plateaus in $\la V_1 \ra$ in Fig.~\ref{fig07} and the
plateau-like features in $\la I_2 - I_3 \ra$ in Fig.~\ref{fig08} using the
following qualitative argument. Since $\la I_2 - I_3 \ra$ mainly gets a
contribution from the Berry curvature term in Eq.~\eqref{ii92}, and
$B_{+,+1,12} \simeq - B_{-,+1,12}$, we can write that equation
approximately as
\beq I_2 - I_3 ~\simeq~ \frac{e}{\hbar} ~4 \hbar \om B_{-,+1,12}
{\dot \phi}_1. \eeq
Let us now replace all the quantities in the above equation by their
average values; this gives
\beq \la I_2 - I_3 \ra ~\simeq~ \frac{e}{\hbar} ~4 \hbar \om {\bar B}_{12}
\la {\dot \phi}_1 \ra. \label{ii93} \eeq
Next, $\la {\dot \phi}_1 \ra$ is related to $\la V_1 \ra$ through
Eq.~\eqref{v1}. On the prominent plateaus in Fig.~\eqref{fig07}, $\la V_1 \ra$
is equal to integer multiples of $\hbar \om/(2e)$. Putting all this together,
Eq.~\eqref{ii93} can be written as
\beq \la I_2 - I_3 \ra ~\simeq~ \frac{e}{\hbar} ~4 n \hbar \om {\bar B}_{12},
\label{ii94} \eeq
where $n$ is an integer. Since we have taken $\hbar \om = 10^{-6}$eV,
Eq.~\eqref{ii94} implies that $\la I_2 - I_3 \ra$ should, in units of
$10^{-6} \times (e/\hbar) ({\rm eV}) \simeq 0.24$ nA, have plateau-like
features at integer multiples of $4 {\bar B}_{12} \simeq 5.36$, where we have
used the value of ${\bar B}_{12}$ quoted after Eq.~\eqref{barb}. This agrees
fairly well with what we observe in Fig.~\ref{fig08}.

Interestingly, the presence of two phases, $\phi_2$ and $\phi_3$,
allows us to vary the widths of the Shapiro plateaus in this
three-wire junction, which would not be possible to do in a two-wire
junction. We have seen in Figs.~\ref{fig07} and \ref{fig09} that the
plateaus are quite wide when $\phi_2 = 2\pi/3$ or zero (recall that
we have fixed $\phi_3 = 0$). However, if we set $\phi_2 = \pi$, we
find from Eq.~\eqref{abs7} that $E_\si$ is independent of $\phi_1$
for any value of $\lam$. The third term on the left hand side of
Eq.~\eqref{ii91} is then equal to zero, and we get a linear equation
in $\phi_1$; hence the Shapiro plateaus disappear completely. Thus
the value of $\phi_2$ can be used to tune the widths of the Shapiro
plateaus.

We have studied plots analogous to Figs.~\ref{fig07} and \ref{fig09} when the
$S$-matrix is of the form given in Eq.~\eqref{srand}, $\hbar \om = 10^{-6}$ eV,
$A = 4$ (in units of $10^{-6} \times (e/\hbar) ({\rm eV}) \simeq 0.24$ nA),
$\al_1 = \al_2 = 0.5$, $\phi_2 =
1.57$, $\phi_3 = 0$, and $\De = 10^{-6}$ eV. We find that the Shapiro plateaus
are very narrow at $\la V_1 \ra$ equal to integer multiples of $\hbar \om/(2e)$
and no plateaus are visible at odd integer multiples of $\hbar \om/(4e)$.
Similarly, no plateau-like features are visible in a plot of $\la I_2 - I_3
\ra$ versus $I$ unlike the plot shown in Fig.~\ref{fig08}.

As discussed at the end of Sec.~\ref{sec:model} E,
the relative widths of the Shapiro plateaus can be understood to some extent
by looking at the absolute values of the Fourier coefficients $|c_l|$ of
$\pa E_{-,+1}/\pa \phi_1$. This is shown in Fig.~\ref{fig10} for three cases:
(a) $S$-matrix given in Eq.~\eqref{smat3} with $\lam=0.1$, $\De=10^{-6}$ eV,
$\phi_2 = 2\pi/3$, and $\phi_3 = 0$, where the Berry curvature has a peak
around $\phi_1 = 4\pi/3$ (see Fig.~\ref{fig03}), (b) $S$-matrix given in
Eq.~\eqref{smat3} with $\lam=0.1$, $\De =10^{-6}$ eV, and $\phi_2 = \phi_3 =
0$, where the Berry curvature has no peak at any value of $\phi_1$
(Fig.~\ref{fig03}), and (c) $S$-matrix given in Eq.~\eqref{srand},
$\De=10^{-6}$ eV, $\phi_2 = 1.57$, and
$\phi_3 = 0$, where the Berry curvature has a peak at $\phi_1 = 2.04$
(Fig.~\ref{fig04}). (In each case, we see that $c_0 =0$ and $|c_{-l}| = |c_l|$
for all values of $l$ as noted after Eq.~\eqref{fourier}). In the system shown
in Fig.~\ref{fig10} (a), we see that $|c_l|$ is quite large at $l$ equal to
both $\pm 1$ and $\pm 2$, although $|c_2/c_1|$ is small; this explains why
there are wide Shapiro plateaus at $\la V_1 \ra$ equal to integer multiples of
$\hbar \om/(2e)$ and narrower plateaus at odd integer multiples of $\hbar
\om/(4e)$ (see Fig.~\ref{fig07}). In Fig.~\ref{fig10} (b), we see that $|c_l|$
is quite large at $l=\pm 1$, but the value of $|c_2/c_1|$ is smaller than in
Fig.~\ref{fig10} (a); this explains why the Shapiro plateaus are wide at
integer multiples $\hbar \om/(2e)$ but quite narrow at odd integer multiples
of $\hbar \om/(4e)$ (Fig.~\ref{fig09}). In Fig.~\ref{fig10} (c), we see that
$|c_l|$ is quite small at all values of $l$; this explains why the Shapiro
plateaus are so narrow in this case.

\begin{widetext}
\begin{center} \begin{figure}
\begin{center}
\subfigure[]{\includegraphics[width=5.9cm]{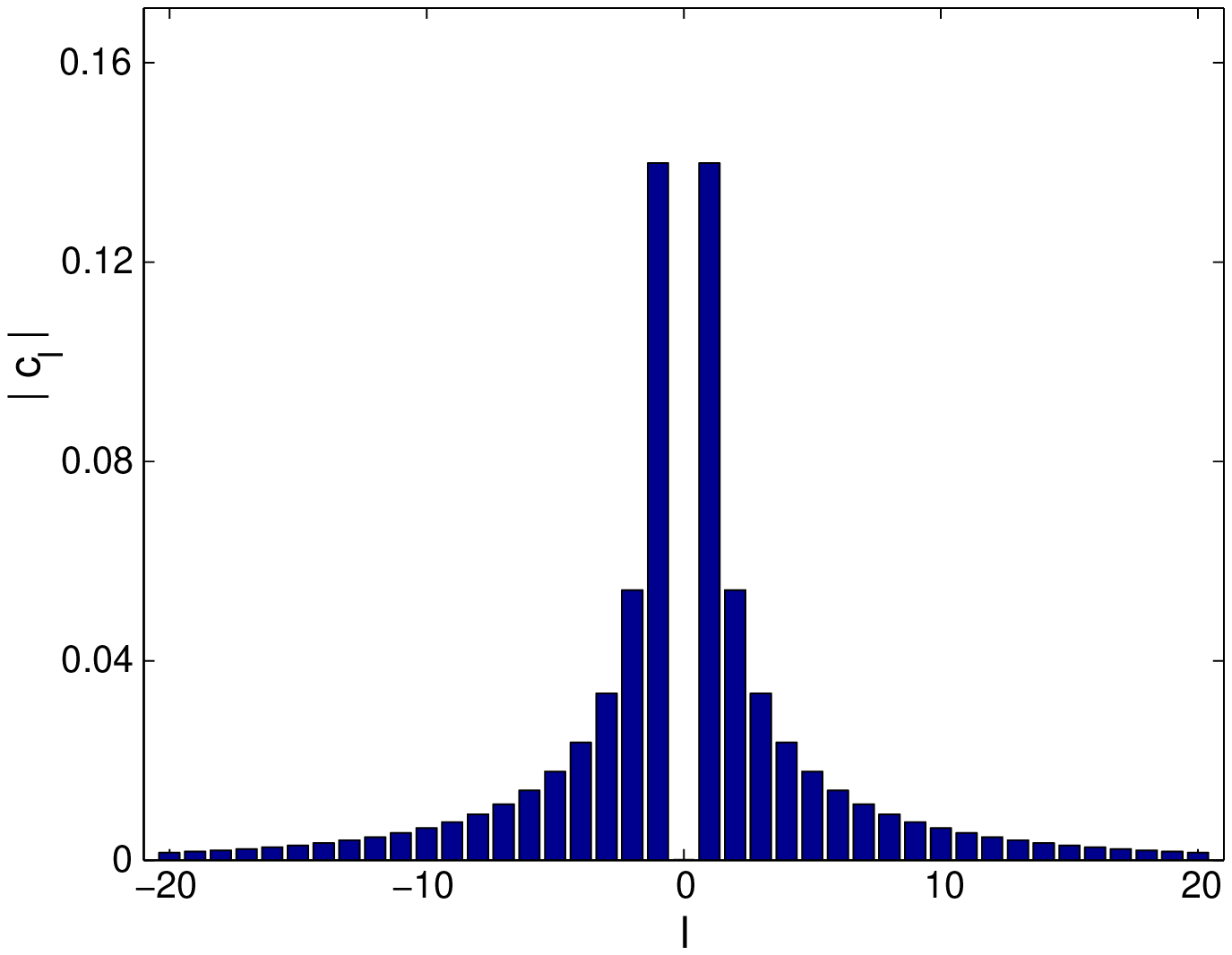}}
\subfigure[]{\includegraphics[width=5.9cm]{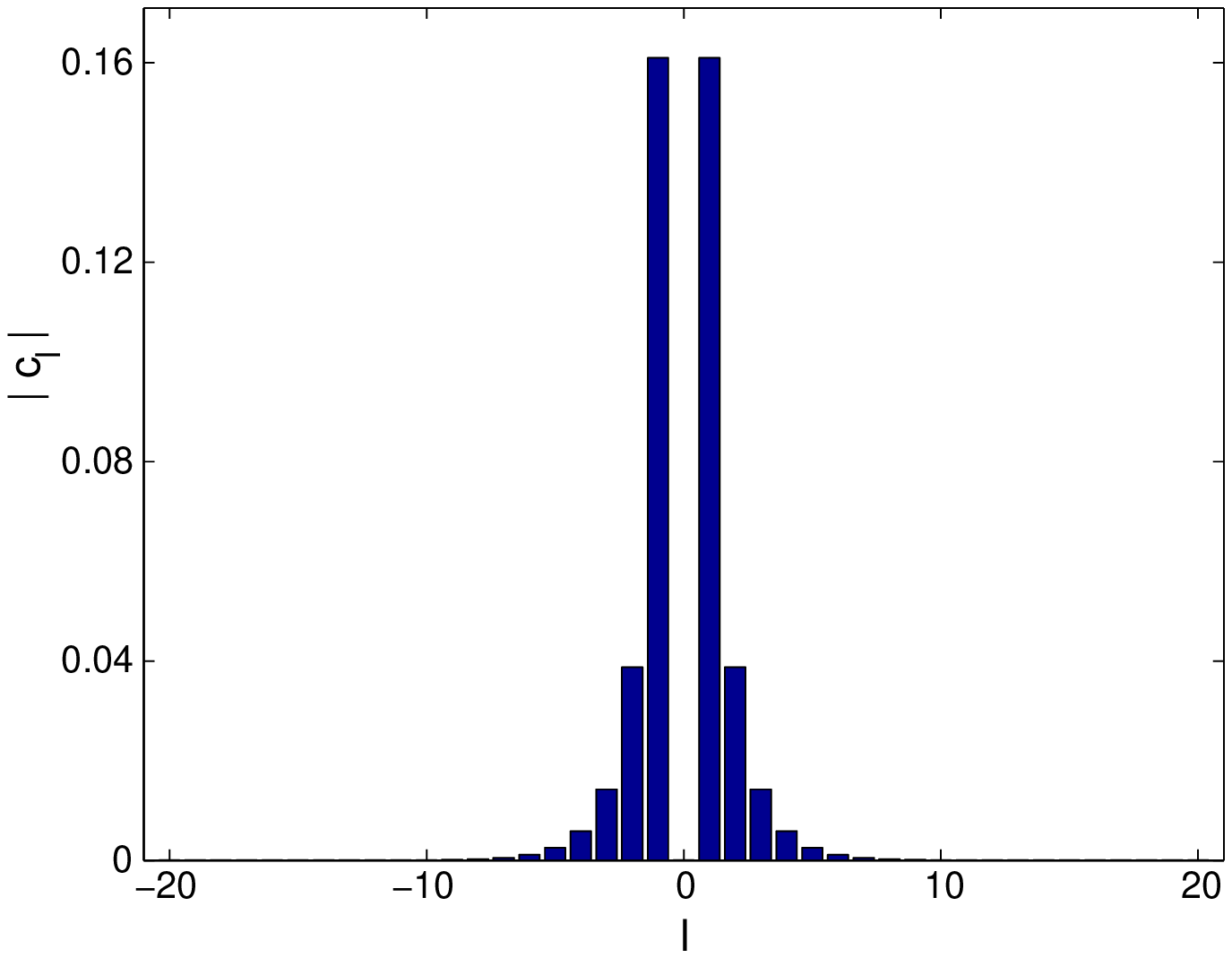}}
\subfigure[]{\includegraphics[width=5.9cm]{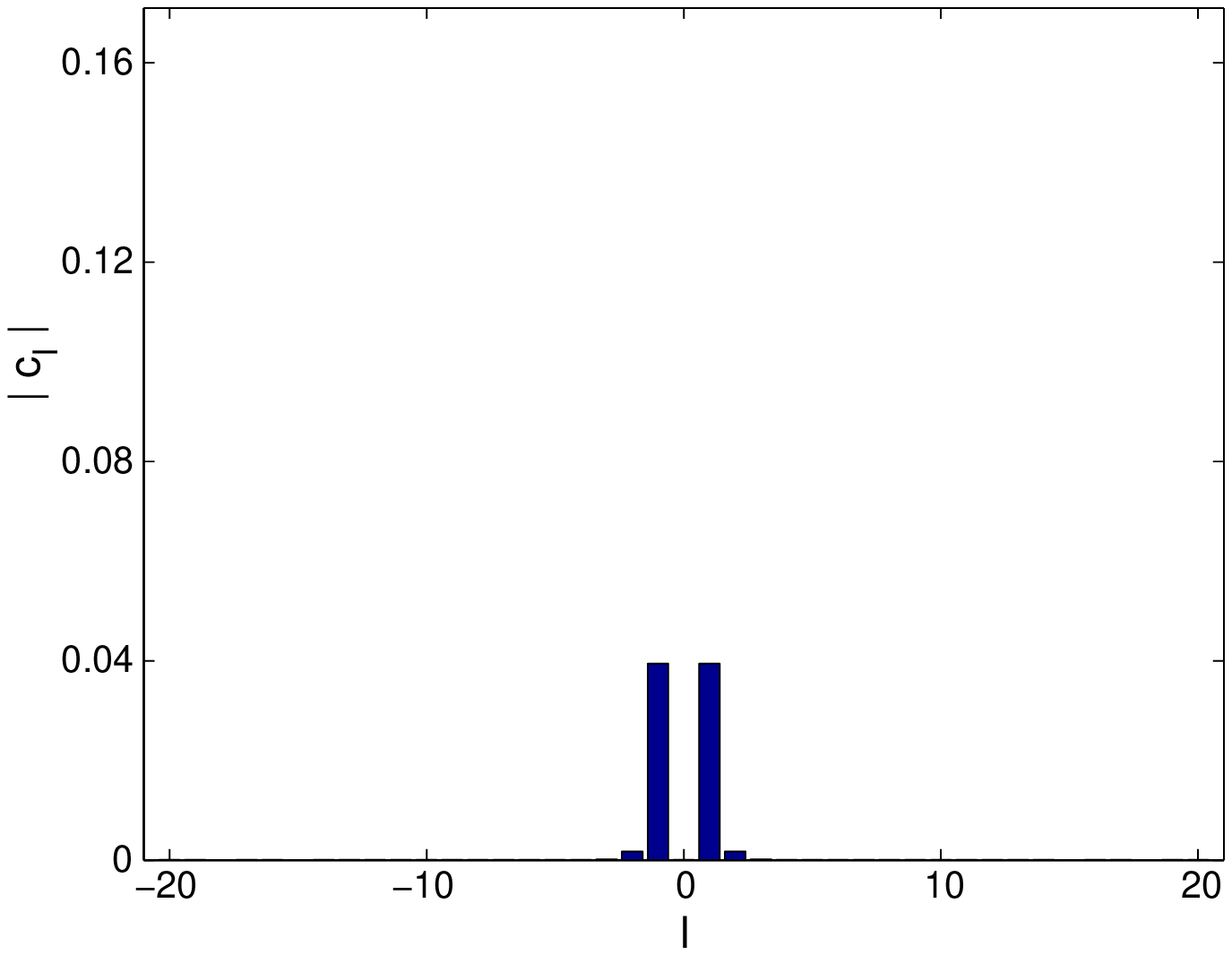}}
\end{center}
\caption{Absolute values of the Fourier coefficients $|c_l|$ (in units of
$10^{-6}$ eV) vs $l$ for three cases: (a) $S$-matrix given in Eq.~\eqref{smat3}
with $\lam=0.1$, $\phi_2 = 2\pi/3$, and $\phi_3 = 0$, (b) $S$-matrix given in
Eq.~\eqref{smat3} with $\lam=0.1$, and $\phi_2 = \phi_3 = 0$, and
(c) $S$-matrix given in Eq.~\eqref{srand}, $\phi_2 = 1.57$, and $\phi_3 = 0$.
We have taken $\De=10^{-6}$ eV in all three cases.} \label{fig10}
\end{figure} \end{center}
\end{widetext}

\begin{figure}[h]
\epsfig{figure=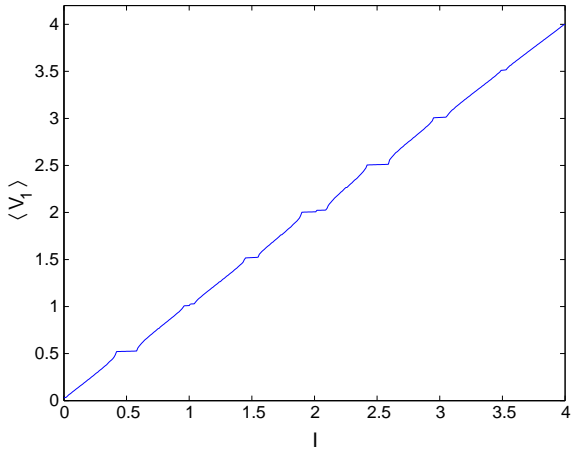,width=8.6cm}
\caption{Plot of $\la V_1 \ra$ vs $I$ for three $p$-wave SC wires. We have
chosen $\hbar \om = 10^{-6}$ eV, $A = 4$, $\al_1 = \al_2 = 0.5$, $\phi_2 =
2\pi/3$, $\phi_3 = 0$, and $\De = 10^{-6}$ eV. ($I$ and $A$ are in units of
$10^{-6} \times (e/\hbar) ({\rm eV}) \simeq 0.24$ nA, and $V_1$ is in units
of $10^{-6}$ eV). The $S$-matrix has the form given in Eq.~\eqref{smat3} with
$\lam=0.1$.} \label{fig11} \end{figure}

\begin{figure}[h]
\epsfig{figure=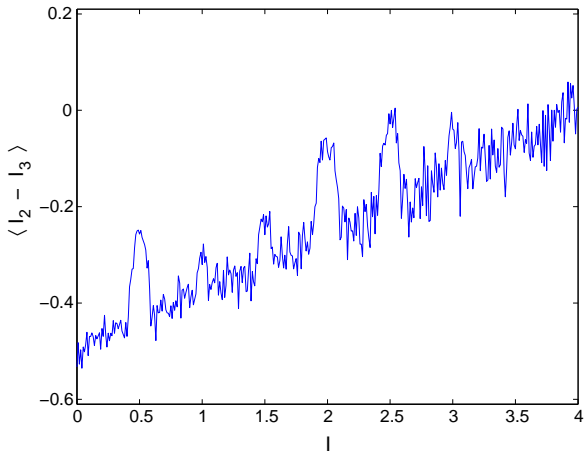,width=8.6cm}
\caption{Plot of $\la I_2 - I_3 \ra$ vs $I$, both in units of $10^{-6} \times
(e/\hbar) ({\rm eV}) \simeq 0.24$ nA. The system parameters are the same
as in Fig.~\ref{fig11}.} \label{fig12} \end{figure}

\subsection{Three $p$-wave superconducting wires}

We have found that Shapiro plateaus also appear in a system with
three $p$-wave superconducting wires. Considering Eqs.~\eqref{abs2}
and \eqref{abs13} for the ABS energies and wave functions for three
$s$-wave and three $p$-wave wires, we find that the band at $E_\si =
\pm \De$ in the first case maps to $E_\si = 0$ in the second case;
the other two energies, given in Eqs.~\eqref{abs5} and
\eqref{abs14}, map from $E_\si/\De > 0 ~(<0)$ in the first case to
$E_\si/\De < 0 ~(>0)$ in the second case; with these mappings, the
corresponding wave functions and Berry curvatures are identical in
the two cases. Figures~\ref{fig11} and \ref{fig12} show $\la V_1
\ra$ and $\la I_2 - I_3 \ra$ versus $I$ when the $S$-matrix has the
form given in Eq.~\eqref{smat3} with $\lam=0.1$, $\hbar \om = 10^{-6}$ eV,
$A = 4$ (in units of $10^{-6} \times (e/\hbar) ({\rm eV}) \simeq 0.24$ nA),
$\al_1 = \al_2 = 0.5$, $\phi_2 = 2\pi/3$, $\phi_3 = 0$, and
$\De = 10^{-6}$ eV. In Fig.~\ref{fig11}, Shapiro plateaus are visible at
integer multiples of $\hbar \om /(2e)$ but not at odd integer
multiples of $\hbar \om /(2e)$. For the ranges of currents in
Fig.~\ref{fig11} where $\la V_1 \ra$ shows plateaus, we see bumps
(rather than plateau-like features) in $\la I_2 - I_3 \ra$ in
Fig.~\ref{fig12}. We find numerically that $\la I_2 - I_3 \ra$ gets
a contribution mainly from the first term (of the form $\pa E/\pa
\phi_i$) in Eq.~\eqref{ii92} which dominates over the Berry
curvature term. Thus, the effects of the Berry curvature are not
easily visible in this system, unlike the case of three $s$-wave wires.

\subsection{Two $p$-wave and one $s$-wave superconducting wire}

We now look at a system in which wires 1 and 2 are $p$-wave SCs and wire
3 is a $s$-wave SC. In this
case, we find that the ABS energies (denoted by $E_{n,\si}$, where $n$
labels the bands and $\si$ labels the spin) are generally not equal to zero or
$\pm \De$. (An exception to this occurs when $\phi_1 = \phi_2$; in this case,
one of the ABS energies is exactly equal to zero). Furthermore, the energies of
the spin-up and -down quasiparticles are generally not equal to each other.
As discussed after Eq.~\eqref{absx2}, we will have $E_{n,\si} = - E_{n,-\si}$
for all values of $(\phi_1,\phi_2)$.

An interesting feature of this system is that one of the ABS energies
can suddenly change from $+\De$ to $-\De$ as we vary the phases $\phi_i$. This
happens whenever the value of one of the $a_\si^2 (E_\si)$'s given by
Eq.~\eqref{abs4u} goes through 1. If we write $a_\si^2 (E_\si) = e^{i2 \ta}$,
we find, using the fact that the imaginary part of $a_\si (E_\si)$ in
Eq.~\eqref{ae}
must be negative or zero, that $E_\si/\De$ is equal to $- \cos \ta$ if $\ta$
is small and positive and is equal to $\cos \ta$ is $\ta$ is small and
negative. Thus, $E_\si$ changes abruptly when $\ta$ goes through zero.

\begin{figure}
\epsfig{figure=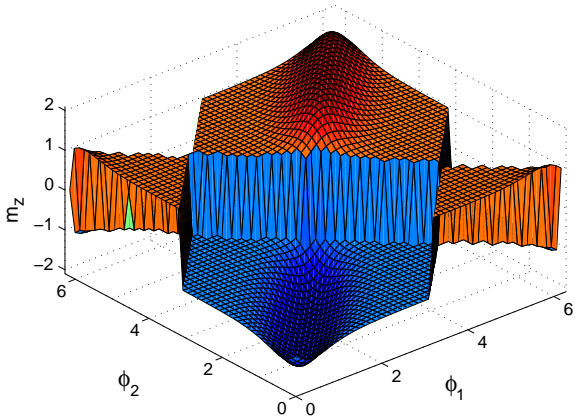,width=8.6cm}
\caption{Surface plot of $m_z$ (in units of $\mu_B/2$) vs $(\phi_1,\phi_2)$
for a system in which superconducting wires 1 and 2 are $p$-wave and wire 3 is
$s$-wave, the $S$-matrix has the form given in Eq.~\eqref{smat3} with
$\lam=0.1$. The temperature has been taken to be $0.1 ~\De /k_B$.}
\label{fig13} \end{figure}

The fact that the ABS energies are not identical for spin-up and and spin-down
quasiparticles implies that there can be some spin-dependent effects in this
system. One such effect is that the region of the junction and the SC wires
can have a net magnetic moment~\cite{sengupta}. [The appearance of
a non-zero magnetic moment requires breaking of time-reversal symmetry. In our
system, this will happen if any of the SC phases $\phi_i$ is not equal to 0 or
$\pi$. This is because $\phi_i \to - \phi_i$ under time-reversal, as we can
see from the complex conjugation of Eq.~\eqref{swavesc} or \eqref{pwavezsc}.
Hence $\phi_i \ne 0$ or $\pi$ breaks time-reversal symmetry. In the numerical
results presented below, we will ensure time-reversal breaking by setting one
of the phases equal to $2\pi/3$.] We define the magnetic moment as
\bea m_z &=& - ~ \frac{\mu_B}{2} ~\sum_{n,\si} ~\si ~[f (E_{n,\si}) ~-~
\frac{1}{2}], \non \\
&=& \frac{\mu_B}{2} ~\sum_n ~\tanh (\beta E_{n,+1}/2), \label{mz} \eea
where $\mu_B$ is the Bohr magneton, the prefactor of $1/2$ has been put
in to avoid double counting of spin, and we have used the symmetry $E_{n,\si}
= - E_{n,-\si}$ to write the second line of Eq.~\eqref{mz}. In
Fig.~\ref{fig13}, we show a surface plot of $m_z$ (in units of $\mu_B/2$)
versus $(\phi_1,\phi_2)$ for a system with the $S$-matrix of the form given in
Eq.~\eqref{smat3} with $\lam=0.1$, and a low temperature given by $T=0.1 ~\De /
k_B$. We have set $\phi_3 = 0$. We see that in large regions of
Fig.~\ref{fig13}, $m_z$ is close to $\pm 1$. This happens
because typically two of the ABS energies $E_{n,+1}$ have the same sign
and the other one has the opposite sign, and $\tanh (z) \to \pm 1$ as $z \to
\pm \infty$. However, near the line $\phi_1 = \phi_2$, one of the energies
gets close to zero and does not contribute much to $m_z$. The other two
energies turn out to have the same sign if $\phi_1 = \phi_2$. Hence the two
contribute with the same sign, either $+1$ or $-1$, producing values of $m_z$
close to $\pm 2$ which is what we observe in Fig.~\ref{fig13}. In the figure,
the vertical faces with long crisscross lines appear because one of the ABS
energies abruptly changes from $+\De$ to $-\De$ as we move across the
$(\phi_1,\phi_2)$ plane, and this leads to an abrupt change in $m_z$. This
abrupt change occurs when either $\phi_1 + \phi_2 = 2\pi$ or $\phi_1 = \phi_2
\pm \pi$.

\begin{figure}
\subfigure[]{\includegraphics[width=8.6cm]{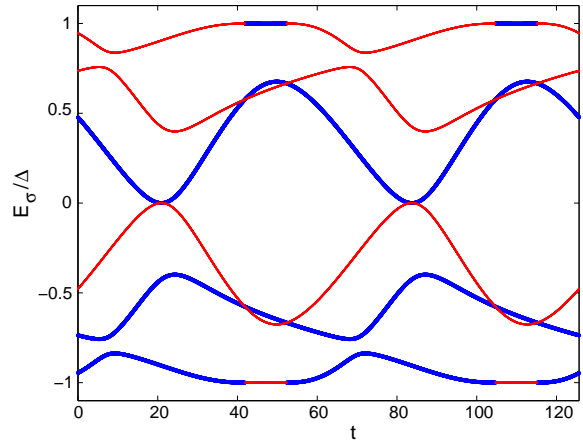}} \\
\subfigure[]{\includegraphics[width=8.6cm]{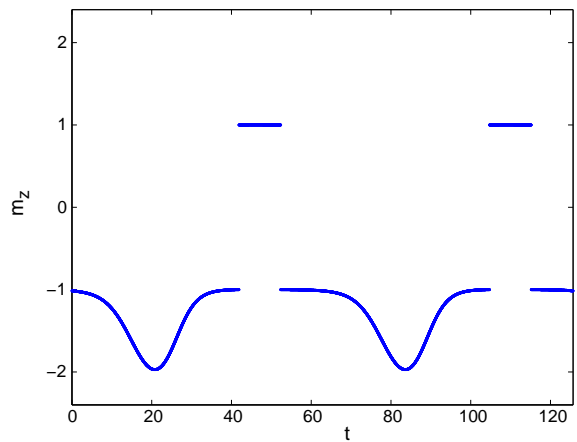}}
\caption{(a) Energies (in units of $\De$) for spin-up (thick blue lines) and
spin-down (thin red lines) quasiparticles vs time (in units of s) for a system
in which superconducting wires 1 and 2 are $p$-wave and wire 3 is $s$-wave, the
$S$-matrix has the form given in Eq.~\eqref{smat3} with $\lam=0.1$, $\phi_2 =
2\pi/3$, $\phi_3 = 0$, and $\phi_1 = 0.1 ~t$. (b) Magnetic moment (in units
of $\mu_B/2$) vs time for the same system at a temperature equal to
$0.1 ~\De /k_B$.} \label{fig14} \end{figure}

\begin{figure}
\subfigure[]{\includegraphics[width=8.6cm]{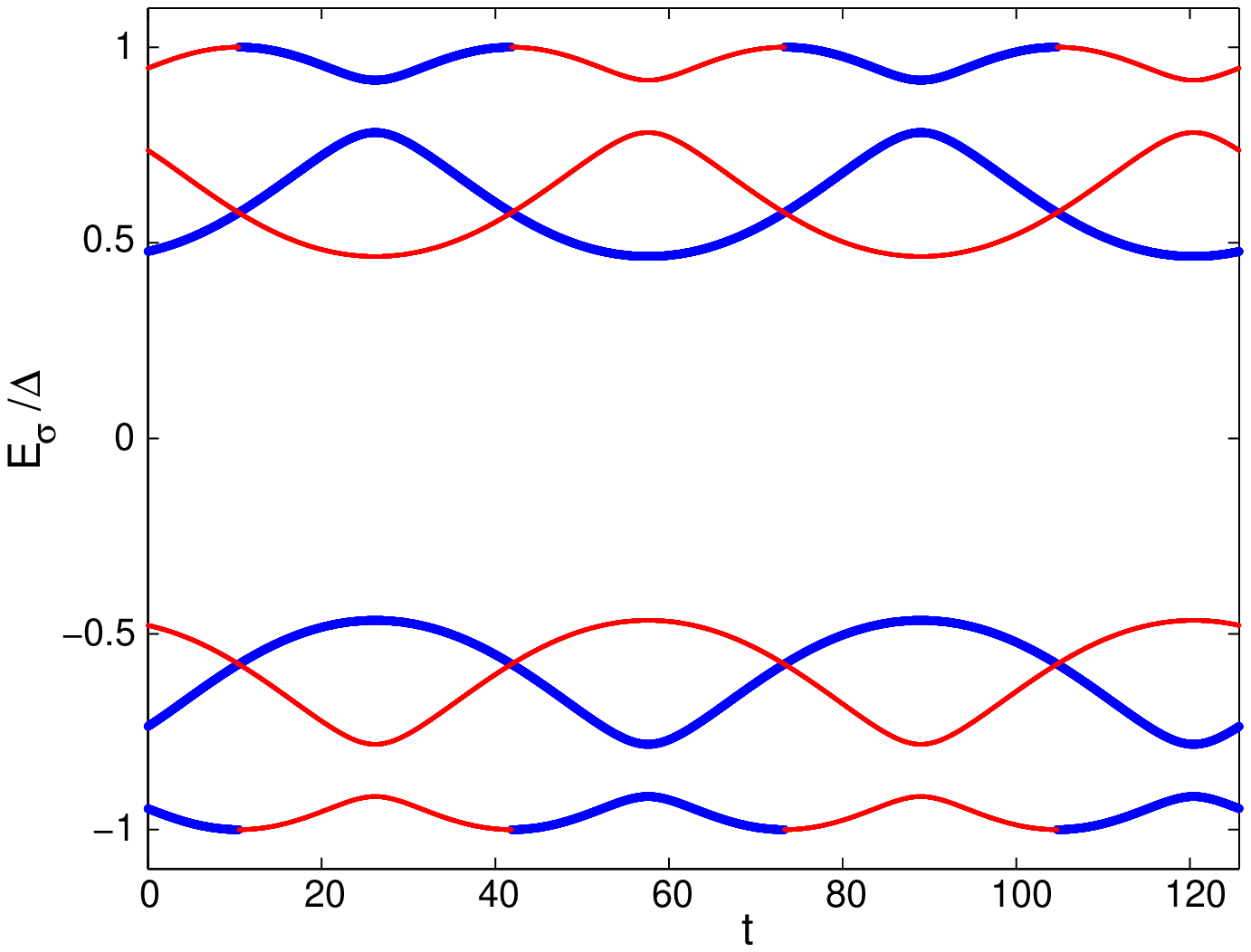}} \\
\subfigure[]{\includegraphics[width=8.6cm]{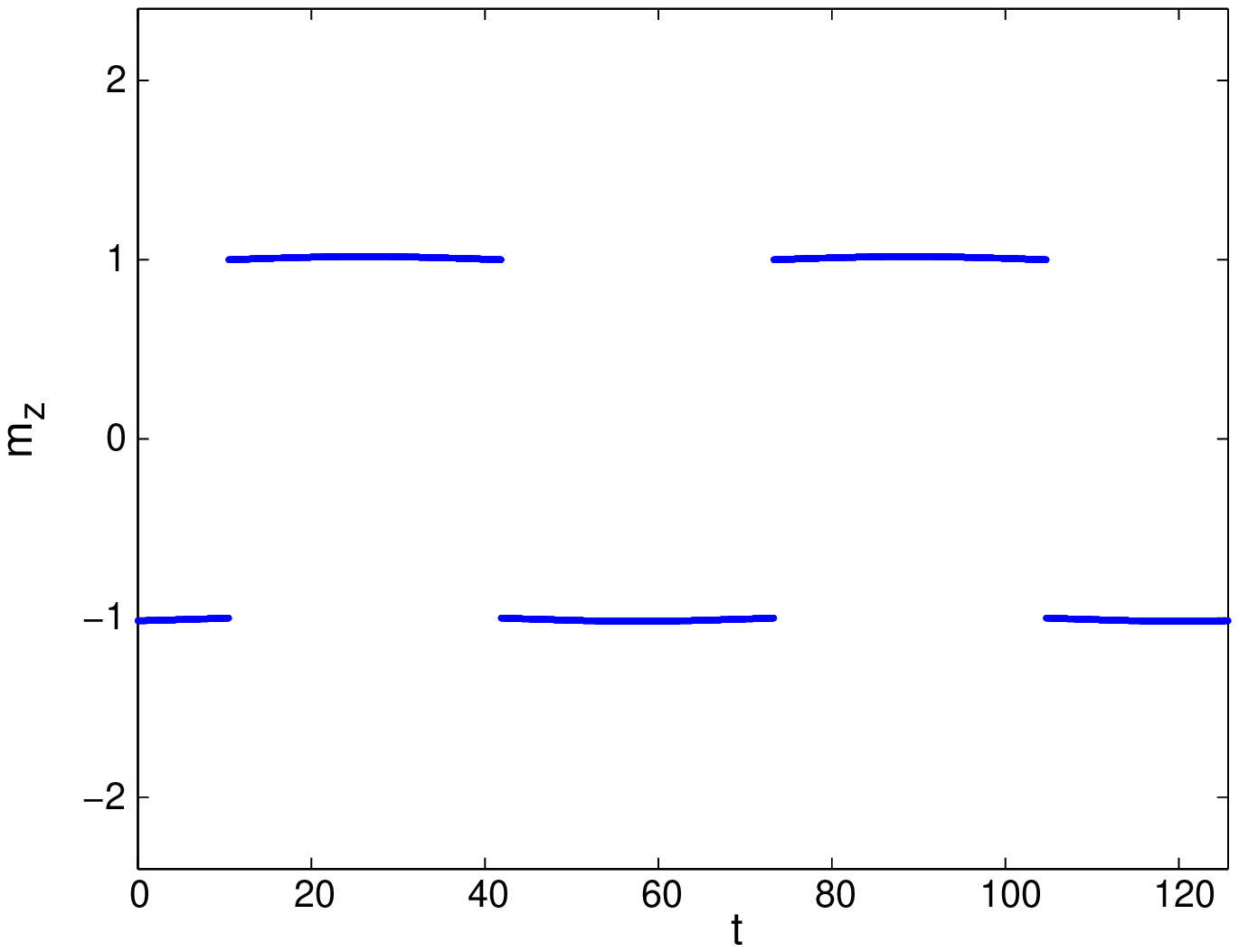}}
\caption{(a) Energies (in units of $\De$) for spin-up (thick blue lines) and
spin-down (thin red lines) quasiparticles vs time (in units of s) for a system
in which superconducting wires 1 and 2 are $p$-wave and wire 3 is $s$-wave, the
$S$-matrix has the form given in Eq.~\eqref{smat3} with $\lam=0.1$, $\phi_1 =
2\pi/3$, $\phi_2 = 0$, and $\phi_3 = 0.1 ~t$. (b) Magnetic moment (in units
of $\mu_B/2$) vs time for the same system at a temperature equal to
$0.1 ~\De /k_B$.} \label{fig15} \end{figure}
\vspace*{.8cm}

Next, we consider what happens when a constant voltage bias $V_1$ is
applied (i.e., to one of the wires with $p$-wave SC)
keeping $V_2 = V_3 = 0$. Then $\phi_1$ varies linearly with time
according to ${\dot \phi}_1 = (2e/\hbar) V_1$ whereas $\phi_2$ and $\phi_3$
remain fixed. As a result, the ABS energies $E_{n,\si}$ vary with time and
hence so does $m_z$. This is shown in Fig.~\ref{fig14} for a system in which
the $S$-matrix is given by Eq.~\eqref{smat3} with $\lam = 0.1$, $\phi_2 =
2\pi/3$, $\phi_3 = 0$, and $\phi_1$ varies with time as $\phi_1 = 0.1 ~t$
(this corresponds to $(2e/\hbar)V_1 = 0.1 ~s^{-1}$ and the initial value
$\phi_1 (t=0) = 0$). Figure~\ref{fig14} shows the results over two time periods
of $\phi_1$, given by $0 \le t \le 4\pi/0.1$. Figure~\ref{fig14} (a) shows the
ABS energies of spin-up (thick blue lines) and spin-down (thin red
lines) quasiparticles as a function of time; we see that one of the energies
for each spin changes suddenly between $+\De$ and $-\De$ at certain times.
Figure~\ref{fig14} (b) shows $m_z$ calculated at the same low temperature equal
to $0.1 ~\De/k_B$ as in Fig.~\ref{fig13}; we see that $m_z$ changes suddenly
between $+1$ and $-1$ at the same times as one of the ABS energies.
When $\phi_1 = \phi_2 = 2\pi/3$, one of the ABS energies for each spin becomes
equal to zero; this happens at $t= (2\pi/3)/0.1 \simeq 20.9$ and $(8\pi/3)/0.1
~\simeq 83.8$. At those times $m_z$ reaches its minimum value of $-2$ as we
see in Fig.~\ref{fig14} (b). (Note that Fig.~\ref{fig14} (b) is essentially a
projection of Fig.~\ref{fig13} on to the line $\phi_2 = 2\pi/3$).

We obtain somewhat different results if a constant voltage bias $V_3$ is
applied (i.e., to the wire with $s$-wave SC) keeping $V_1 = V_2 = 0$. Then
$\phi_3$ varies linearly with time according to ${\dot \phi}_3 = (2e/\hbar)
V_1$ whereas $\phi_1$ and $\phi_2$ remain fixed. The variations of the ABS
energies $E_{n,\si}$ and $m_z$ with time are shown in Fig.~\ref{fig15} over
two time periods for a
system in which the $S$-matrix is given by Eq.~\eqref{smat3} with $\lam = 0.1$,
$\phi_1 = 2\pi/3$, $\phi_2 = 0$, and $\phi_3$ varies with time as $\phi_3 =
0.1 ~t$. Figure~\ref{fig15} (a) shows the ABS energies of spin-up (thick blue
lines) and spin-down (thin red lines) quasiparticles as a function of time; we
again see that one of the energies for each spin changes suddenly between
$+\De$ and $-\De$ at certain times. Figure~\ref{fig15} (b) shows $m_z$
calculated at a temperature equal to $0.1 ~\De/k_B$; we see that $m_z$ changes
suddenly between $+1$ and $-1$ at the same times as one of the ABS energies.
This occurs when $\phi_1 + \phi_2 = 2\phi_3$ mod $2\pi$.
Note that none of the ABS energies ever become equal to zero since
$\phi_1 \ne \phi_2$; hence $m_z$ does not become equal to $\pm 2$ at any time.

\section{Discussion}
\label{sec:discussion}

In this paper, we have studied a system of several SC wires which
meet at a junction. The junction is parametrized by a scattering
matrix $S$ which is of a normal form that does not mix electrons and
holes. The SC wires can have $s$-wave or $p$-wave pairings, with the
pairing phases denoted by $\phi_i$. We have discussed how the ABS
energies and wave functions can be determined for any combination of
$s$-wave and $p$-wave wires and any spin of the quasiparticles.
Our results provide a scattering matrix based formalism for
studying junctions involving superconducting wires with both $s$- and
$p$-wave pairings; these results generalize the earlier existing results
for multi-terminal $s$-wave wires~\cite{yoko,riwar}. Various
symmetries of the energies and wave functions have been pointed out.
In particular, spin-up and -down ABS have the same energies if
the three wires are of the same type (all $s$-wave or all $p$-wave),
but they do not have the same energies if some of the wires are $s$-wave
and others are $p$-wave. We have then studied the Berry curvature and Chern
numbers which appear if the number of SC wires is three or more.

Next, we have discussed the ac Josephson effect in which a constant
voltage bias is applied to one of the SC wires. We find that an
time-averaged current flows only in the other two wires; this
current is sensitive to the Berry curvature and the
corresponding integrated transconductance is quantized in units of
$4e^2/h$. We then discuss what happens if resistances and
capacitance are placed between every pair of wires and if the
current flowing in one of the wires has both a constant piece $I$
and a piece which is sinusoidally varying with a frequency $\om$. In
such an $RC$ circuit Shapiro plateaus can appear in a plot of the
time-averaged voltage bias $\la V_1 \ra$ in the same
wire versus $I$. The plateaus can occur at values of $\la V_1 \ra$
equal to any rational multiple of $\hbar \om /(2e)$
leading, in principle, to a devil's staircase structure. However, in
practice, the plateau width goes to zero rapidly as the denominator
of the rational number becomes large. We have shown that the plateau
widths are related to Fourier transforms of the ABS energies as
functions of the $\phi_i$'s.

Next, we have studied several systems of three SC wires in detail. As two
examples of the scattering matrix $S$, we have considered a highly symmetric
and time-reversal invariant form and a randomly generated asymmetric form
which is not time-reversal invariant. [Throughout our analysis, we have
assumed that the matrix $S$ is spin independent even when time-reversal
symmetry is broken. This would be true if, for example, the junction consists
of a loop which is threaded by a magnetic flux (thus breaking time-reversal
symmetry); such a flux would affect the transport around the loop through 
an Aharonov-Bohm phase which does not depend on the spin~\cite{meyer}.]
We present explicit expressions for the ABS energies for the cases of three 
$s$-wave and three $p$-wave wires. The forms of the Berry curvature and the
values of the Chern numbers depend crucially on the form of $S$.

Finally, we have numerically studied a number of three-wire systems
to test the various ideas presented in the earlier sections. For
three $s$-wave wires, we have shown that the dependence of the ac
Josephson current in one wire on the phase $\phi_i$ in one of the
other wires can directly provide information about the Berry
curvature. Next, we have shown that when the time-averaged $V$ in
one wire is plotted versus $I$ when the current also contains an
oscillating piece in the same wire, Shapiro plateaus with
discernible widths appear at both integer and half-odd-integer
multiples of $\hbar \om /(2e)$. Furthermore, the currents in the
other two wires show plateau-like features in the ranges of the $I$
where $V$ shows plateaus; these features are mainly due to the Berry
curvature terms in the currents. For three $p$-wave wires, we find
Shapiro plateaus in the plot of $V$ versus $I$; the currents in the
other wires show some bumps in the same ranges of $I$ but these are
not primarily due to the Berry curvature. For a system with two
$p$-wave wires and one $s$-wave wire, the fact that the ABS energies
are not identical for spin-up and -down ABS leads to
spin-dependent effects. For instance, we find that the junction
region can have a non-zero magnetic moment which depends on the
phases $\phi_i$. We note that this allows for a direct control
of the time variation of such a magnetic moment by externally applying
a voltage across the wires. In the ac Josephson effect, where one of
these phases varies linearly with time, we find that the magnetic
moment varies periodically in time, showing large jumps when
one of the ABS energies either touches zero or changes
abruptly between the top and the bottom of the SC gap.

There are several experiments which could verify our theoretical
results. First, we predict that standard $I-V$ characteristics
measurements with three-wire $s$-wave or $p$-wave junctions in the
presence of an external microwave radiation of frequency $\om$
would detect Shapiro plateaus at rational fractional values of $2e
\la V_1 \ra/(\hbar \om)$. We note that this is in contrast to
two-wire junctions where such plateaus can be seen in standard
experiments at integral values $2eV_1/(\hbar \om)$. Such
experiments are routine for two-wire junctions~\cite{exp1,exp2} and
could, in principle, be carried out for multi-wire systems. Second,
for three-wire junctions involving both $p$- and $s$-wave wires, one
may detect the effect of an oscillating magnetic moment through the
resulting radiated electric field near the junction; such
experiments have been discussed in a different context~\cite{kanda}.
Finally, the measurement of the integrated transconductance in
junctions involving three $s$-wave wires by using an appropriate
four-terminal setup would constitute an experimental way of
ascertaining the quantization predicted by Eq.~\eqref{i2c12}.

In conclusion, we have studied multi-wire junctions of $s$- and $p$-
wave superconductors and have developed a scattering matrix based
approach which can be used to describe such junctions. We have
studied the ac Josephson effect in such junctions pointing out the
presence of Shapiro plateaus. We have also shown that such systems
involving both $s$- and $p$-wave superconducting wires lead to
presence of magnetic moments in the junction whose time variation
can be controlled via an external applied voltage. We have suggested
several experiments which can test our theory.

\vspace*{.5cm} \section*{Acknowledgments}

D.S. thanks Department of Science and Technology, India for Project No.
SR/S2/JCB-44/2010 for financial support.

\end{document}